\documentclass[12pt]{article}

\usepackage[margin=1in]{geometry}
\usepackage{hyperref}
\usepackage{mathrsfs}
\usepackage{enumitem}
\usepackage{amsmath}
\usepackage{amssymb}
\usepackage{mathtools}
\usepackage{amsthm}
\usepackage{float}
\usepackage{tabularx}
\usepackage{algorithm}
\usepackage{makecell}
\usepackage{algpseudocode}
\usepackage{microtype}
\usepackage{graphicx}
\usepackage{authblk}
\usepackage{array}
\usepackage{adjustbox}
\usepackage{subcaption}
\usepackage{longtable}
\usepackage{bm}
\usepackage[table]{xcolor}
\usepackage{booktabs}
\usepackage{placeins}
\usepackage[capitalize,noabbrev]{cleveref}

\usepackage{tikz}
\usetikzlibrary{calc}

\usepackage{siunitx}
\newcommand{\wtl}[3]{$#1\,/\,#2\,/\,#3$}

\theoremstyle{plain}
\newtheorem{theorem}{Theorem}[section]

\theoremstyle{definition}
\newtheorem{definition}[theorem]{Definition}

\theoremstyle{remark}

\algrenewcommand\algorithmicrequire{\textbf{Input:}}
\algrenewcommand\algorithmicensure{\textbf{Output:}}

\usepackage{titlesec}
\titleformat{\section}{\normalfont\bfseries\large}{\thesection}{1em}{}
\titleformat{\subsection}{\normalfont\bfseries}{\thesubsection}{1em}{}

\title{\bfseries Learning Fast Monomial Orders for Gr\"obner Basis Computations}

\author[1]{R. Caleb Bunch}
\author[2]{Alperen A. Erg\"ur}
\author[2]{Melika Golestani}
\author[3]{Jessie Tong}
\author[4]{Malia Walewski}
\author[5]{Yunus E. Zeytuncu}

\affil[1]{Georgia Institute of Technology}
\affil[2]{The University of Texas at San Antonio}
\affil[3]{San Diego State University}
\affil[4]{Emory University}
\affil[5]{University of Michigan--Dearborn}

\date{}

\begin{document}
\maketitle

\begin{abstract}
\noindent
The efficiency of Gr\"obner basis computation, the standard engine for solving systems of polynomial equations, depends on the choice of monomial ordering. Despite a near-continuum of possible monomial orders, most implementations rely on static heuristics such as GrevLex, guided primarily by expert intuition. We address this gap by casting the selection of monomial orderings as a reinforcement learning problem over the space of admissible orderings. Our approach leverages domain-informed reward signals that accurately reflect the computational cost of Gr\"obner basis computations and admits efficient Monte Carlo estimation. Experiments on benchmark problems from systems biology and computer vision show that the resulting learned policies consistently outperform standard heuristics, yielding substantial reductions in computational cost. Moreover, we find that these policies resist distillation into simple interpretable models, providing empirical evidence that deep reinforcement learning allows the agents to exploit non-linear geometric structure beyond the scope of traditional heuristics.

\end{abstract}

\section{Introduction}
Systems of non-linear algebraic equations arise naturally in  kinematics \cite{nielsen1999kinematics,wampler2011kinematics,raghavan1995kinematics}, cryptography \cite{courtois2000efficientcrypto,bard2009algebraiccrypto,park2018sidecrypto}, systems biology \cite{perez2018structurebio,feinberg2019foundationsbio,dickenstein2020algebraicbio}, multiple view geometry \cite{hartley2004multiplegeo, nister2004efficientgeo,agarwal2024atlasgeo},   statistical inference \cite{diaconis1998algebraicstats,sullivant2018algebraicstats,almendra2024markovstats}, and discrete optimization \cite{de2012algebraicdiscrete, cohen2013algebraicdiscrete, conti1991buchbergerdiscrete}.  
To transform this  descriptive power into prescriptive capacity, one must either solve the underlying system of polynomial equations or gain insight on their structure. This is where the challenge lies. Quantifying the intrinsic difficulty of computing the common zeros of polynomial systems precisely is an important ongoing discussion in post-quantum cryptography and complexity theory. We also note that for certain structured problems, alternatives to Gr\"obner bases may exist. For example, \cite{Larsson} shows that in specific computer vision settings, non-Gr\"obner bases can lead to faster minimal solvers. Notwithstanding these discussions, we focus on the most general-purpose and widely applicable method for solving polynomial systems: Gr\"obner basis computation.

A Gr\"obner basis can be viewed as a far-reaching generalization of Euclid’s algorithm, extending it from the world of integers to the realm of multivariate polynomial systems and discrete geometry. Remarkably, this extension was formalized more than two millennia after Euclid, with the introduction of Gr\"obner bases in Buchberger’s thesis \cite{Buchberger2006}. Buchberger also proposed an algorithm for computing such bases, and despite substantial advances over the intervening decades, state-of-the-art methods remain largely refinements of this original framework \cite{Faugere1999F4,faugere2002F5}. These methods are best understood not as fully specified algorithms, but as algorithmic schemes that require the user to make several critical choices. While these choices do not affect theoretical correctness, they can have a dramatic impact on practical computational performance.

To instantiate a Buchberger-style algorithm, two principal decisions must be made: the choice of a monomial ordering and the strategy for selecting pairs for reduction. We carefully define these two notions in the next section. In practice, most implementations default to a small set of standard monomial orderings, such as Lex, GrLex, and GrevLex, despite the existence of a near-continuum of valid alternatives. More precisely, the space of all admissible monomial orderings is encoded by a rich geometric object known as the Gr\"obner fan. This structure  is largely ignored in current implementations, which instead rely on a handful of default choices derived from expert intuition. In this work, we challenge this convention by casting the selection of monomial orderings as a reinforcement learning problem.

\subsection{Our contributions}
Our primary contribution demonstrates the potential of reinforcement learning (RL) as a tool for computational algebra by directly challenging the conventional wisdom that GrevLex should serve as the default monomial ordering. Across polynomial systems arising in systems biology and computer vision, our trained agents consistently outperform standard default choices, achieving reductions in computational cost of up to 70\% under a rigorously calibrated proxy. We also provide direct comparisons to classic black-box optimization heuristics, namely Random Search and Simulated Annealing, establishing that while reinforcement learning excels on several systems, there remain very large search spaces where classical search methods are still competitive.

We focus our training and data generation on zero-dimensional polynomial systems, that is, systems with finitely many common solutions. This choice is strategic. For such systems, computing a Gr\"obner basis with a fast, agent-selected ordering enables efficient conversion to the lexicographical (Lex) ordering using the FGLM algorithm and its modern variants. This capability is essential in applications requiring variable elimination, where direct computation in Lex order is often infeasible. Moreover, across diverse applications of Gr\"obner basis in sciences and engineering  zero-dimensional systems are the ones that appear most of time. This motivates our focus on zero dimensional ideals.

We design a reinforcement learning environment in which agents learn to select optimized monomial orderings. To the best of our knowledge, this represents the first application of RL to this fundamental and widely used problem in computational algebra. The environment incorporates reward signals informed by domain expertise and is constructed to allow efficient estimation of the reward. We empirically validate the reward design by comparing it against measured runtimes, observing that it closely tracks computational complexity. We expect that this reward formulation and environment design will be useful beyond the present work, supporting future research at the intersection of reinforcement learning and symbolic computation. We found that the scientific computing libraries in Python only allowed few fixed orderings; as a result we used Julia and wrote the RL environment from scratch. Our code and data are available \href{https://github.com/maliawalewski/umich-reu-2025}{here}.

Finally, we investigate whether the learned policies can be distilled into simpler, interpretable models. Using a synthetic dataset of over 100 problems, we train agents and attempt distillation into symbolic expressions and soft decision trees of varying complexity. These experiments suggest that the agents rely on a relatively small subset of exponent vectors when determining effective monomial orderings, indicating that careful subsampling may serve as a useful heuristic prior to optimization. At the same time, the distilled models fail to capture the full behavior of the learned policies. This provides empirical evidence that the agents exploit the subtle geometric structure of the Gr\"obner fan that is not easily represented by the basic interpretable models considered here. We outline potential directions for developing richer interpretable models in the concluding section.

\subsection{Related work}
The literature on Gr\"obner bases and its connections to algebraic geometry is extensive. Applications of Gr\"obner basis also form a vast literature. Rather than attempting a broad survey, we focus on prior work at the intersection of machine learning and Gr\"obner basis computation, which is most closely aligned with the aims of this paper.

A notable recent research direction, initiated by \cite{kambe2024learning, kambe2025geometric}, employs transformer models to predict Gr\"obner traces. While this line of work has shown promising results, it addresses a complementary problem. Our focus is not on predicting traces produced by a fixed algorithm, but on selecting monomial orderings that induce favorable traces for Buchberger’s algorithm. Other studies have applied machine learning to related symbolic computation tasks, including predicting the computational complexity of Gr\"obner basis computations \cite{jamshidi2023predicting} and optimizing elimination schemes in cylindrical algebraic decomposition \cite{huangengland2014applying, huangengland2016using}. These works represent independent directions from our problem in this paper.

The work most closely related to ours is the ICML 2020 paper by Peifer, Stillman, and Halpern-Leistner \cite{peiferstillman2020}, which applies reinforcement learning to optimize pair selection strategies in Buchberger’s algorithm. In their formulation, the learning environment changes substantially after each action, leading to a highly non-stationary setting that is computationally demanding and constrains large-scale experimentation. In contrast, we formulate monomial ordering selection as a stationary and comparatively simple reinforcement learning environment. This design choice is motivated by two considerations: reducing the computational cost of training and evaluation, and for  preserving "clarity of direction"  towards the goals of learning. The latter is inspired by recent evidence that relatively simple reinforcement learning schemes, such as Group Relative Policy Optimization (GRPO) \cite{shao2024deepseekmath}, can be very effective when the underlying signal is well aligned with the task structure.

\section{A Focused Introduction to  Gr\"obner Bases and Buchberger's Algorithm}
In this section, we briefly review key mathematical concepts and notation; for further details see \cite{CoxLittleOShea2015}. Let \(k\) be a field and \(R = k[x_1,\dots,x_n]\) the polynomial ring over \(k\). Fix a monomial ordering \(\prec\) on \(R\). For a nonzero polynomial \(f \in R\), denote by \(\mathrm{LM}_\prec(f)\) and \(\mathrm{LT}_\prec(f)\) its leading monomial and leading term, respectively.

\begin{definition}
A finite set \(G = \{g_1,\dots,g_t\} \subset I \subset R\) is a \emph{Gr\"obner basis} of \(I\) with respect to \(\prec\) if
\[
\langle \mathrm{LT}_\prec(g_1), \dots, \mathrm{LT}_\prec(g_t) \rangle
=
\langle \mathrm{LT}_\prec(I) \rangle.
\]
\end{definition}
A Gr\"obner basis is \emph{reduced} if each \(g_i\) is monic and no monomial of \(g_i\) is divisible by \(\mathrm{LM}_\prec(g_j)\) for \(j \neq i\). For a fixed monomial order, every ideal admits a unique reduced Gr\"obner basis \cite{CoxLittleOShea2015}. This does not imply, however, that every order will create a different reduced Gr\"obner basis as there are complex equivalences encoded by the Gr\"obner fan.

\subsection{Monomial Orderings and Weight Vectors}

A monomial ordering is a well-ordering on monomials that is compatible with multiplication. This means for every collection of monomials there will be a smallest one, and for any two monomials $x^{\alpha} =x_1^{\alpha_1} x_2^{\alpha_2} \cdots x_n^{\alpha_n}$ and $x^{\beta}=x_1^{\beta_1} \cdots x_n^{\beta_n}$ if $x^{\alpha}$ is ordered greater than $x^{\beta}$ then $x^{\alpha} x^{\gamma}= x^{\alpha+\gamma}$ should be ordered greater than $x^{\beta} x^{\gamma} = x^{\beta + \gamma}$. These properties ensure we can perform division and elimination properly, and guarantee
the termination of Gr\"obner basis algorithms. There is a convenient way to introduce monomial orders: Weight vectors \(w=(w_1,\dots,w_n)\in\mathbb{R}_{>0}^n\) can be used to create orderings via
\[
\deg_w(x^\alpha)=\sum_{i=1}^n w_i\alpha_i,
\]
with ties broken by a secondary order. Allowing real weights is natural when studying Gr\"obner fans, since this ordering is invariant under multiplying all coordinates of the weight vector with a fixed scalar. We mentioned that most current implementations only allow a few fixed monomial orders derived from expert intuition. These orderings are 
\begin{itemize}
    \item \textbf{Lex}: Lexicographic ordering starting from $x_1$ to $x_n$ (a limit of \(w=(M^{n-1},\dots,1)\), \(M\gg1\)),
    \item \textbf{GrLex}: This orders first with total degree and then breaks ties with lexicographic ordering (corresponding to \(w=(1,\dots,1)\)).
    \item \textbf{GrevLex}: This orders first with total degree and then breaks ties with reverse-lexicographic ordering (corresponding to \(w=(1,\dots,1)\)).
\end{itemize}
The choice of ordering strongly affects both the structure and the computational cost of the resulting Gr\"obner basis. There are theoretical and empirical results suggesting GrevLex is efficient, however, no general method exists for selecting an optimal order \cite{CoxLittleOShea2015}: this is precisely the blind spot in conventional wisdom that this paper aims to clarify.

\subsection{Buchberger's Algorithm and S-Polynomials}

Buchberger’s algorithm \cite{Buchberger2006} is based on \emph{S-polynomials}.

\begin{definition}
For nonzero \(f,g\in R\),
\[
S(f,g)=
\frac{\mathrm{lcm}(\mathrm{LM}(f),\mathrm{LM}(g))}{\mathrm{LT}(f)}f
-
\frac{\mathrm{lcm}(\mathrm{LM}(f),\mathrm{LM}(g))}{\mathrm{LT}(g)}g.
\]
\end{definition}

A generating set \(G\) is a Gr\"obner basis if and only if all S-polynomials reduce to zero modulo \(G\). Although conceptually simple, Buchberger’s algorithm can exhibit severe intermediate expression growth and doubly exponential worst-case complexity \cite{mayr1982complexity}. For graded reverse lexicographic orderings, sharper single-exponential degree bounds are known \cite{lazard1983,dube1990}, motivating their widespread use.

\subsection{The Gr\"obner Fan}

For a fixed ideal \(I \subset k[x_1,\dots,x_n]\), different monomial orders may yield distinct reduced Gr\"obner bases and initial ideals. The \emph{Gr\"obner fan} \cite{MoraRobbiano1988GrobnerFan,FukJensenThomas2007} provides a geometric framework for organizing this dependence using weight vectors \(w \in \mathbb{R}^n\). For \(f=\sum_\alpha c_\alpha x^\alpha\), the \(w\)-initial form \(\mathrm{in}_w(f)\) consists of terms with maximal \(\langle w,\alpha\rangle\), and the associated initial ideal is
\[
\mathrm{in}_w(I)=\langle \mathrm{in}_w(f): f\in I\rangle.
\]

As \(w\) varies, \(\mathrm{in}_w(I)\) remains constant on relatively open polyhedral cones and changes only when ties occur. These cones form a polyhedral fan whose full-dimensional cones are in bijection with the reduced Gr\"obner bases of \(I\). Classical orders such as Lex, GrLex, and GrevLex correspond to specific directions or refinements within this fan. Figure~\ref{fig:grobner-fan-example} illustrates this structure schematically on a two-dimensional slice; it conveys how regions of constant initial ideal arise and how crossing a cone boundary corresponds to a combinatorial change in leading terms.

\medskip
\noindent\textbf{Example.}
Following Example~2.7 of \cite{FukJensenThomas2007}, consider
\[
I=\langle x+y+z,\; x^3z+x+y^2\rangle \subset \mathbb{Q}[x,y,z].
\]
With lexicographic order \(x\prec y\prec z\), a reduced Gr\"obner basis is
\[
G_{\mathrm{lex}}(I)=\{z+y+x,\; y^2+x-x^3y-x^4\},
\]
so \(\mathrm{in}_{\mathrm{lex}}(I)=\langle z,y^2\rangle\). The corresponding Gr\"obner cone consists of weight vectors satisfying linear inequalities ensuring that these leading terms persist; when one of these inequalities becomes tight, the initial ideal changes.

\begin{figure}[t]
\centering
\begin{tikzpicture}[scale=5, line cap=round, line join=round]
  \coordinate (X) at (0,0);
  \coordinate (Y) at (1,0);
  \coordinate (Z) at (0.5,0.8660254);

  \coordinate (C)  at (0.5,0.2886751);
  \coordinate (P)  at (0.7142857,0.3711537);

  \draw[thick] (X)--(Y)--(Z)--cycle;
  \fill[gray!10] (X)--(Y)--(Z)--cycle;

  \node[below left]  at (X) {$w_1$};
  \node[below right] at (Y) {$w_2$};
  \node[above]       at (Z) {$w_3$};

  \draw[thick] (C) -- ($(X)!0.48!(Z)$);
  \draw[thick] (C) -- ($(Y)!0.48!(Z)$);
  \draw[thick] (C) -- ($(X)!0.48!(Y)$);

  \fill (C) circle(0.0045);
  \node[xshift=-1pt,yshift=11pt] at (C)
   {\scriptsize $(\tfrac13,\tfrac13,\tfrac13)$};

  \fill (P) circle(0.0045);
  \node[xshift=4pt,yshift=-9pt] at (P)
    {\scriptsize $(\tfrac17,\tfrac37,\tfrac37)$};

  \node at (0.28,0.12)
    {\scriptsize $\mathrm{in}_w(I)=\langle z,y^2\rangle$};
\end{tikzpicture}
\caption{A schematic two-dimensional slice ($w_1+w_2+w_3=1$, $w\ge0$) of the Gr\"obner fan for
\(I=\langle x+y+z,\;x^3z+x+y^2\rangle\), illustrating how regions of constant
\(\mathrm{in}_w(I)\) arise. The diagram is not to scale and is intended to convey
the geometric structure of cones rather than exact fan boundaries; see
\cite{FukJensenThomas2007}.}
\label{fig:grobner-fan-example}
\end{figure}

\subsection{The FGLM and \(F_4\) Algorithms}

For zero-dimensional ideals \(I\), where \(R/I\) is finite-dimensional, the FGLM algorithm \cite{FGLM1993} converts a Gr\"obner basis under a degree-compatible order into one under another order using linear algebra. Modern variants are commonly used to transform from a fast monomial ordering to Lex, and the computation is effective in applications; see \cite{FaugereMou2017SparseFGLM,berthomieu2022faster}.

Faug\`ere’s \(F_4\) algorithm \cite{Faugere1999F4} accelerates Buchberger’s method by batching reductions and performing them via structured linear algebra. At each stage, selected S-polynomials and multiples of basis elements are assembled into a matrix whose rows correspond to polynomials and columns to monomials. A subroutine of \(F_4\) called symbolic preprocessing constructs this matrix, after which reduction is performed by Gaussian elimination. \(F_4\) exploits the specific sparsity patterns of these matrices for high performance. 

The dominant cost of \(F_4\) is governed by the dimensions of these matrices. In particular, the number of columns, denoted by \(n_M\) in our reward formulation, captures the size of the active monomial basis and directly determines memory usage and linear algebra cost. Degree growth increases these dimensions and depends strongly on the monomial order. Since computation proceeds incrementally through such matrix constructions, these quantities provide natural proxies for computational effort and are used as feedback signals in our RL environment design.

\vspace{-0.2 in}
\section{Reinforcement Learning Setup}
\subsection{Problem Formulation and Environment Design}
The problem we study is highly sensitive to changes in the support set (the collection of all exponent vectors present in the equations). In contrast, it exhibits significantly less sensitivity to variations in the coefficients of the defining equations, provided the monomials remain fixed. This observation was known from computer vision applications, where coefficients vary across problem instances while the support set remains constant; consequently, the trace Gr\"obner basis computation remains relatively stable. We leverage this stability in our environment design. The reinforcement learning environment consists of thousands of problem instances, and the reward is estimated by sampling these instances and computing the average reward.

\subsection{Computational Framework}
Common Python-facing computational algebra tooling we considered does not expose the weighted monomial order search space and trace statistics we required. To address this, we utilized the Julia package \texttt{Groebner.jl} \cite{demin2024groebnerjl}. Our decision to use \texttt{Groebner.jl} was motivated primarily by its accessibility and efficiency. 
\begin{enumerate}
    \item \textbf{Accessibility:} It is highly approachable for researchers outside the computational algebra community.
    \item  \textbf{Efficiency:} It implements the F4 algorithm, which is an efficient yet relatively straightforward variant of Buchberger's method.
\end{enumerate} Since the Julia reinforcement learning ecosystem is not as mature as Python's, we developed our RL environment from scratch. Our code is available at the following \href{https://github.com/maliawalewski/umich-reu-2025}{repository}.

\subsection{Reward Function}

\begin{figure}[tb]
  \centering
  \includegraphics[
    width=0.70\linewidth,
    height=0.33\textheight,
    keepaspectratio
  ]{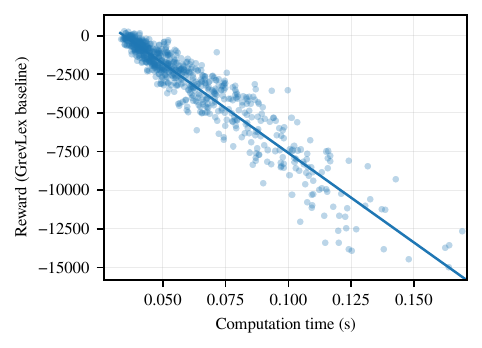}
  \caption{Reward improvement (relative to GrevLex) versus runtime for Gr\"obner basis computation on a fixed randomly generated ideal. Each point corresponds to a weighted monomial ordering with weights \((i,j,k)\in\{30,35,\ldots,70\}^3\) ($N=729$).}
  \label{fig:rewardvstime}
\end{figure}

Our reward function utilizes the Gr\"obner trace statistics of the F4 algorithm recorded by \texttt{Groebner.jl}, specifically:  the number of iterations ($t$), the number of columns in the elimination matrix ($n_M(i)$) for $1 \leq i \leq t$, the number of pairs ($n_P(i)$), and the degree of pairs ($d_P(i)$). The reward function is defined as:
\[ - \sum_{i=1}^t n_M(i) n_P(i) \ln(d_P(i))  \]
The intuition behind this formulation is that the degree contributes to evaluation complexity on a logarithmic scale ($\ln(d)$), while the primary discrete parameters governing cost are the pair count and the size of the matrix. 

We validated this reward by sweeping 729 weighted orderings on a fixed randomly generated ideal and comparing reward improvement over GrevLex against the measured runtime of computing a Gr\"obner Basis. The relationship is strongly monotone (Pearson $r=-0.949$, Spearman $\rho=-0.937$; see Figure ~\ref{fig:rewardvstime}). In our implementation we used the expert intuition, GrevLex, as a base-line for our rewards. This has two justifications: using a good baseline reduces the noise in training, and we have the easy interpretation that the agent is rewarded only when doing better than GrevLex and punished otherwise.

\subsection{Action Space and Weight Vectors}
The \texttt{Groebner.jl} package accepts only integer entries in weight vectors for the monomial ordering. However, the theoretical search space is the simplex: 
\[ \Delta_n := \{ x=(x_1,x_2,\ldots,x_n)  : x_i \geq 0, \sum x_i =1 \} \]
Therefore, we define the action space as the simplex and define a monomial ordering by scaling the action vector by $10^3$ and rounding to the nearest integer.

\subsection{Agent Interaction and Training}
At each step during an episode, the agent outputs a weight vector that defines a weighted monomial order. Using this order, we run a Gr\"obner basis computation and obtain a reward given by a Monte Carlo approximation of the reward using sample instances. The agent then updates the weight vector and repeats. An episode consists of a fixed number of such updates and terminates after \(L\) steps. We treat the episode length \(L\) and the number of training episodes \(N\) as hyper-parameters. Rewards are averaged over a batch of \(B=10\) ideals per step, where each ideal is instantiated from the same support set and uses uniform random coefficients from the finite field \(\mathbb{F}_{32003} = \mathbb{Z}/32003\mathbb{Z}\). In experiments on ideals arising in systems biology and computer vision, we found \(N=10{,}000\) episodes and \(L=25\) steps per episode to work well.

An important practical detail is that a weight vector coordinate approaching zero effectively ignores a variable. This represents a degenerate choice from an algebraic perspective and introduces instability in practical computations. Consequently, we experimented with shrinking the action space by enforcing a lower bound on every coordinate, requiring $x_i \geq n^{-\alpha}$ for various integer values of $\alpha$. 

\subsection{Model Architecture}
Due to the sensitivity of the problem to input changes and our limited computational resources, we utilized the TD3 architecture with a prioritized replay buffer for its stability and efficiency \cite{fujimotoTD3,schaul2015PER}. We experimented with various activation functions to address vanishing gradients including GELU (Gaussian Error Linear Unit) \cite{hendrycks2016gelu}, but found that ReLU worked well in most cases. See Table~\ref{tab:nn-arch} in Appendix~\ref{app::experimental-details} for the neural network architecture used in the experiments. Additionally, we tested different architectures with memory, such as an LSTM for the actor component of TD3, but observed only marginal performance improvements. We suspect this may be due to the current scale of our ideals, and further large-scale experimentation may provide deeper insights. Finally, needless to say that like most RL implementations, we performed extensive hyperparameter tuning to optimize the performance (see Appendix~\ref{app::experimental-details}). 

\section{Experiments on Problems from Systems Biology and Computer Vision}

\begin{table}[t]
\centering
\scriptsize
\setlength{\tabcolsep}{3pt}
\renewcommand{\arraystretch}{1.1}
\caption{RL agent test performance (pooled over 5 seeds). Win, Tie, Loss mean percentages are reported along with mean percent reward improvement conditioned on wins and mean percent reward degradation conditioned on losses. Testing was done over \(500{,}000\) test ideals (\(100{,}000\) per seed).}
\label{tab:main-results-reward}
\begin{tabular}{l
                c S[table-format=2.2] S[table-format=3.2]
                c S[table-format=2.2] S[table-format=3.2]}
\toprule
& \multicolumn{3}{c}{\textbf{GrevLex baseline}} & \multicolumn{3}{c}{\textbf{GrLex baseline}} \\
\cmidrule(lr){2-4}\cmidrule(lr){5-7}
\textbf{Polynomial System} &
\textbf{Win/Tie/Loss} &
{\textbf{Improvement}} &
{\textbf{Degradation}} &
\textbf{Win/Tie/Loss} &
{\textbf{Improvement}} &
{\textbf{Degradation}} \\
\midrule
Relative Pose &
\wtl{98.4}{0.0}{1.6} & 19.43 & -1.27 &
\wtl{32.2}{16.1}{51.6} & 7.95 & -7.04 \\
Triangulation &
\wtl{27.0}{66.1}{7.0} & 0.86 & -0.72 &
\wtl{100}{0}{0} & 37.62 & {--} \\
$n$-Site Phos.\ ($n{=}14$) &
\wtl{100}{0}{0} & 70.77 & {--} &
\wtl{100}{0}{0} & 70.77 & {--} \\
Wnt Shuttle &
\wtl{91.6}{0.0}{8.4} & 54.64 & -26.40 &
\wtl{86.9}{0.0}{13.1} & 55.25 & -20.83 \\
\bottomrule
\end{tabular}
\end{table}

We tested the RL model on several important polynomial systems arising in computer vision, including minimal problems in relative pose with unknown focal length \cite{stewenius2008minimal} and three-view triangulation \cite{stewenius2005howhard}, as well as models from systems biology such as the \(n\)-site phosphorylation system \cite{giaroli2019parameterregions} and the Wnt shuttle model \cite{gross2016algebraicwnt}. See Table \ref{tab:main-results-reward} for the main results. 

The RL models were all trained on Google Cloud Compute Engine instances. For the relative pose problem, three-view triangulation, and \(n\)-site phosphorylation (\(n=14\)) we used a \textit{c4-highcpu-8} instance (8 vCPUs, 16 GB Memory). For the Wnt shuttle model system we used a \textit{c4-highmem-8} instance (8 vCPUs, 62 GB Memory) due to the size of the polynomial support set. See Appendix \ref{app::experimental-details} for experimental details. The summary of results are presented in \Cref{tab:main-results-reward}. We report the percentage of instances where the agents outperform standard heuristics during testing, the percentage of improvement in those cases, and the percentage of degradation when agents were beaten by expert heuristics. We should note that relative pose, triangulation, and \(n\)-site phosphorylation are polynomial systems in 3, 3, and 2 variables respectively, while the Wnt shuttle model yields a system in 19 variables. This is reflected in the RL training curves, with more variables leading to a larger search space and greater variance during training. 

\subsection{Relative Pose with Unknown Focal Length}
This benchmark is the two-view relative pose problem where both cameras are calibrated
up to a \emph{shared unknown focal length} (with otherwise fixed intrinsics such as centered
principal point, unit aspect ratio, and zero skew) \cite{stewenius2008minimal}.
Given six point correspondences (the minimal case), one estimates the epipolar geometry
and the focal parameter by enforcing the epipolar constraint together with the essential-matrix
constraints induced by the unknown focal length \cite{stewenius2008minimal}.
The resulting constraints yield a polynomial system that can be expressed in the linear
form $MX=0$, where $M\in\mathbb{R}^{10\times 33}$ and $X$ is a vector of monomials in the
chosen unknowns \cite{stewenius2008minimal}. In our experiments we order the unknowns as
$(l_1,l_2,p)$ and use the associated monomial support listed in Appendix~\ref{app:relpose-support}.

After training, the RL agent was evaluated on \(500{,}000\) test ideals pooled over five random seeds. As shown in Table~\ref{tab:main-results-reward}, the learned ordering substantially outperforms GrevLex on the relative pose system, winning in \(98.4\%\) of test instances with a mean reward improvement of \(19.4\%\) when it wins, and only rare degradations (\(1.6\%\) losses with a mean degradation of \(1.3\%\)). In contrast, performance relative to GrLex is mixed, with the agent winning in \(32.2\%\) of cases and losing in \(51.6\%\), reflecting the strong suitability of GrLex for this problem. These results indicate that the RL agent reliably discovers orderings competitive with or superior to standard heuristics, particularly compared to GrevLex. The final learned weight vectors across seeds are reported in Appendix \ref{app::relative-pose-additional-results} (see Table \ref{tab:relative-pose-final-weights}).

\subsection{Three-View Triangulation}
Optimal three-view triangulation seeks a 3D point $X\in\mathbb{P}^3$ (homogeneous coordinates)
minimizing the sum of squared reprojection errors in three calibrated views:
\[
C(X) \;=\; \sum_{i=1}^3 d(P_iX, x_i)^2,
\]
where $P_i\in\mathbb{R}^{3\times 4}$ are known projection matrices and $x_i$ are observed image points
\cite{stewenius2005howhard}. The stationary conditions $\partial C/\partial X_j=0$ ($j=1,2,3,4$)
can be cleared of denominators to obtain polynomial equations in the homogeneous coordinates
\cite{stewenius2005howhard}. Using the homogeneity of $C$, one reduces to three simultaneous
degree-6 polynomial equations in the inhomogeneous variables $(x,y,z)$ (after fixing scale by setting $t=1$) \cite{stewenius2005howhard}. We order variables as $(x,y,z)$ and use the supports listed in
Appendix~\ref{app:triangulation-support}.

After training, the RL agent was evaluated on \(500{,}000\) test ideals pooled over five random seeds. As shown in Table~\ref{tab:main-results-reward}, the learned ordering performs comparably to GrevLex on the three-view triangulation system, with a majority of test instances resulting in ties (\(66.1\%\)) and only modest gains when the agent wins (mean improvement \(0.86\%\)). We should note that triangulation is seed-sensitive: one seed attains near-universal wins against GrevLex, while the remaining seeds are dominated by ties (Appendix~\ref{app::triangulation-additional-results}). In contrast, the agent consistently outperforms GrLex, winning on all test instances with a mean reward improvement of \(37.6\%\). This pattern indicates that GrevLex already induces a near-optimal ordering for this problem, while the RL agent reliably avoids the substantial inefficiencies associated with GrLex. Overall, the learned ordering matches the strongest classical heuristic while remaining robust across baseline choices. The final learned variable weight vectors across seeds are located in Appendix \ref{app::triangulation-additional-results} (see Table \ref{tab:triangulation-final-weights}).

\subsection{$n$-Site Phosphorylation System}
This benchmark comes from a sequential $n$-site phosphorylation/dephosphorylation mechanism
modeled with mass-action kinetics and conservation laws \cite{giaroli2019parameterregions}.
For the class of subnetworks considered in \cite{giaroli2019parameterregions}, the phosphatase total
is constant, so substituting a (positive) monomial parametrization of steady states reduces the
conservation relations to a system of \emph{two} equations in the two unknowns
$(s_0,e)$ (unphosphorylated substrate and kinase) \cite{giaroli2019parameterregions}.
Equivalently, the reduced steady-state equations can be written in sparse matrix form $CX=0$ with
$C\in\mathbb{R}^{2\times(n+3)}$ and $X$ a vector of the $(n+3)$ monomials appearing in the two equations
\cite{giaroli2019parameterregions}. In our experiments we set $n=14$ (so $|X|=17$) and order variables
as $(s_0,e)$, using the shared support listed in Appendix~\ref{app:nsite-support}.

After training, the RL agent was evaluated on \(500{,}000\) test ideals pooled over five random seeds. As shown in Table~\ref{tab:main-results-reward}, the learned ordering consistently outperforms both GrevLex and GrLex on the \(n\)-site phosphorylation system, winning on all test instances against both baselines and achieving a large mean reward improvement of \(70.8\%\). No degradations were observed in either comparison. These results indicate that the RL agent reliably discovers highly effective monomial orderings for this class of sparse biochemical systems, substantially improving upon standard heuristics. The final variable weight vectors across seeds can be found in Appendix \ref{app::n-site-additional-results} (see Table \ref{tab:nsite-final-weights}).

\subsection{Wnt Shuttle Model}
The Wnt shuttle model captures protein shuttling between cellular compartments and the regulation
and degradation of $\beta$-catenin within the Wnt signaling pathway \cite{maclean2015parameterfree}.
Mathematically, it is formulated as a mass-action reaction network with 19 species (variables
$x_1,\dots,x_{19}$) and 31 reaction-rate parameters $k_1,\dots,k_{31}$ \cite{gross2016algebraicwnt}.
The dynamics admit five independent linear conservation relations, introducing conserved totals
$c_1,\dots,c_5$ \cite{gross2016algebraicwnt}. Steady states are characterized by the 19 polynomial
equations $\dot x_i=0$ together with the five conservation laws, yielding a polynomial system of
24 equations in the 19 species variables (with parameters appearing in coefficients)
\cite{gross2016algebraicwnt}. We use the monomial supports of these polynomials as listed in
Appendix~\ref{app:wnt-support} (see \cref{eq:wnt-support-set}).

After training, the RL agent was evaluated on \(500{,}000\) test ideals pooled over five random seeds. As shown in Table~\ref{tab:main-results-reward}, the learned ordering substantially outperforms both GrevLex and GrLex on the Wnt shuttle system, winning in \(91.6\%\) and \(86.9\%\) of test instances, respectively. When the agent improves upon the baseline, the gains are large, with mean reward improvements exceeding \(54\%\) in both comparisons. However, the agent also exhibits non-negligible losses relative to both heuristics, reflecting the complexity of this large polynomial system. These results demonstrate that the RL agent frequently identifies orderings that significantly accelerate Gr\"obner basis computation on the Wnt shuttle model, while occasional degradations remain. The final variable weight vectors across seeds are reported in Appendix \ref{app::wnt-additional-results} (see Table \ref{tab:wnt-final-weights}).

\begin{figure}[tb]
  \centering
  \includegraphics[
    width=0.70\linewidth,
    height=0.33\textheight,
    keepaspectratio
  ]{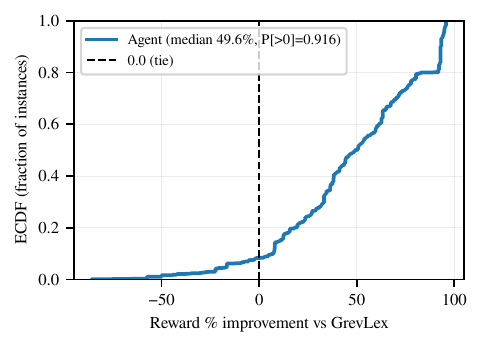}
  \caption{WNT shuttle model: empirical CDF of per-instance test-time reward percent improvement relative to GrevLex, $\Delta r_{\%} = 100\,(r_{\text{agent}} - r_{\text{GrevLex}})/(|r_{\text{GrevLex}}|+\varepsilon)$ (pooled over 5 seeds; $10^5$ test ideals per seed). The dashed vertical line marks $\Delta r_{\%}=0$ (tie); values to the right indicate instances where the agent achieves higher reward than GrevLex. The median improvement is $49.6\%$ and $\Pr[\Delta r_{\%}>0]=0.916$.}
  \label{fig:wnt-ecdf}
\end{figure}

\subsection{Benching Against Baseline Methods} \label{sec:baselines}

To better validate the performance of our reinforcement learning agents, we conducted direct comparisons against two classical black-box optimization baselines: Random Search (RS) and Simulated Annealing (SA). These methods were chosen for their robustness and ability to navigate combinatorial spaces without requiring gradient information. 

We evaluated all three methods across the four base sets (Relative Pose, Triangulation, $n$-Site Phosphorylation, and Wnt Shuttle Model). Each method was evaluated over 5 independent seeds, with agents and baselines tested on the same distribution of $10^5$ polynomial systems per seed (yielding $5 \times 10^5$ total test instances per base set). Table~\ref{tab:baseline-comparison} summarizes the win rates, mean improvements, and degradations for RS, SA, and TD3 against both the default GrevLex and GrLex orderings. 

\begin{table}[htb!]
\centering
\scriptsize
\setlength{\tabcolsep}{3pt}
\renewcommand{\arraystretch}{1.1}
\caption{Comparison of Random Search (RS), Simulated Annealing (SA), and Reinforcement Learning (TD3) against the GrevLex and GrLex baselines. Metrics are pooled over $5 \times 10^5$ test instances per problem class.}
\label{tab:baseline-comparison}
\begin{tabular}{ll
                c S[table-format=2.2] S[table-format=3.2]
                c S[table-format=2.2] S[table-format=3.2]}
\toprule
& & \multicolumn{3}{c}{\textbf{GrevLex baseline}} & \multicolumn{3}{c}{\textbf{GrLex baseline}} \\
\cmidrule(lr){3-5}\cmidrule(lr){6-8}
\textbf{Polynomial System} & \textbf{Method} &
\textbf{Win/Tie/Loss} &
{\textbf{Imp.}} &
{\textbf{Deg.}} &
\textbf{Win/Tie/Loss} &
{\textbf{Imp.}} &
{\textbf{Deg.}} \\
\midrule
Relative Pose & RS & \wtl{54.4}{0.0}{45.6} & 27.32 & -19.57 & \wtl{51.6}{0.0}{48.4} & 12.21 & -44.80 \\
              & SA & \wtl{96.6}{0.0}{3.4}  & 17.23 & -5.54  & \wtl{49.0}{0.0}{51.0} & 6.68 & -11.49 \\
              & TD3& \wtl{98.4}{0.0}{1.6}  & 19.43 & -1.27  & \wtl{32.2}{16.1}{51.6} & 7.95 & -7.04 \\
\midrule
Triangulation & RS & \wtl{100.0}{0.0}{0.0} & 0.91 & -3.40 & \wtl{100.0}{0.0}{0.0} & 38.10 & {--} \\
              & SA & \wtl{100.0}{0.0}{0.0} & 0.91 & -2.10 & \wtl{100.0}{0.0}{0.0} & 38.10 & {--} \\
              & TD3& \wtl{27.0}{66.1}{7.0} & 0.86 & -0.72 & \wtl{100.0}{0.0}{0.0} & 37.62 & {--} \\
\midrule
$n$-Site Phos.& RS & \wtl{100.0}{0.0}{0.0} & 20.30 & {--} & \wtl{100.0}{0.0}{0.0} & 20.30 & {--} \\
              & SA & \wtl{100.0}{0.0}{0.0} & 70.73 & {--} & \wtl{100.0}{0.0}{0.0} & 70.73 & {--} \\
              & TD3& \wtl{100.0}{0.0}{0.0} & 70.77 & {--} & \wtl{100.0}{0.0}{0.0} & 70.77 & {--} \\
\midrule
Wnt Shuttle   & RS & \wtl{100.0}{0.0}{0.0} & 57.96 & {--} & \wtl{100.0}{0.0}{0.0} & 57.44 & {--} \\
              & SA & \wtl{100.0}{0.0}{0.0} & 93.42 & {--} & \wtl{100.0}{0.0}{0.0} & 93.47 & {--} \\
              & TD3& \wtl{91.6}{0.0}{8.4} & 54.64 & -26.40 & \wtl{86.9}{0.0}{13.1} & 55.25 & -20.83 \\
\bottomrule
\end{tabular}
\end{table}

As shown in Table~\ref{tab:baseline-comparison}, the TD3 agent matches or outperforms the classical baselines on the Relative Pose, Triangulation, and $n$-Site Phosphorylation systems against GrevLex. In these settings, RL effectively captures the underlying geometric structure of the Gr\"obner fan to rapidly identify superior monomial orderings. However, on the highly complex 19-variable Wnt Shuttle Model, the classical baselines (particularly Simulated Annealing) achieve higher win rates and greater mean improvements. The Wnt system's massive 19-dimensional search space poses a significant exploration challenge for the RL agent, causing it to occasionally get lost in suboptimal regions where heuristic methods are able to maintain more consistent performance. This highlights an important trade-off: while RL is highly effective on many algebraic systems, scaling it to very large combinatorial spaces remains an open challenge where traditional black-box heuristics still provide strong, reliable baselines.

Crucially, however, the reinforcement learning agent is significantly more sample-efficient. As detailed by the training curves in Appendix~\ref{app::additional-results}, the TD3 agent typically converges to a highly performant policy within the first \(1{,}000\) episodes across most systems. In contrast, Simulated Annealing relies on high-variance stochastic exploration and often requires thousands of episodes to recover from initial degradations and stabilize at a comparable reward level.

\section{Policy Distillation}

To better understand the successful strategies of the agents, we generated a synthetic dataset comprising over 100 polynomial systems where the RL agents outperformed default heuristics (see Table \ref{tab:synthetic-results}).  We attempted to distill these learned strategies into interpretable formats using two methods: soft decision trees and symbolic equations.  Soft decision trees provide a trainable and interpretable model and has been used in policy distillation successfully, see \cite{gokhale2024distill2explain}. Fitting symbolic equations is the canonical method in basic sciences and has seen a current surge of interest in interpretable machine learning in general and for policy distillation in particular \cite{cranmer2020discovering,li2025symbolic}.  

\begin{table}[tb]
\centering
\small
\caption{Average RL agent reward performance against GrevLex and GrLex over synthetic ideals.}
\label{tab:synthetic-results}
\begin{tabular}{lcc}
\toprule
\textbf{RL Agent vs Baseline} & \textbf{Win \%} & \textbf{Improvement \%} \\
\midrule
RL Agent vs GrLex   & 99.04 & 16.70 \\
RL Agent vs GrevLex & 94.78 & 11.70 \\
\bottomrule
\end{tabular}
\end{table}

\subsection{Symbolic Regression}
We used \texttt{SymbolicRegression.jl} \cite{cranmer2023interpretable} to search for simple closed-form equations mapping polynomial ideal support sets to monomial ordering weight vectors. Following the default MLJ selection rule in \texttt{SymbolicRegression.jl}, we selected a single expression from the discovered Pareto frontier for each target as the highest-scoring expression among those with loss at most \(1.5\) times the minimum loss on the frontier (see Appendix \ref{app::symbolic-best-equations}). We trained separate symbolic regression models to predict each entry of the weight vector (for \(n=3\) variables, this yields 3 equations). The selected equations were then fixed and evaluated by comparing the monomial order induced by the predicted variable weights against GrevLex.

\subsection{Complexity of the Gr\"obner Fan}
While the distilled decision trees achieved low Mean Squared Error (MSE) on training data, they exhibited limited generalization on previously unseen test sets. Similarly, the best-fit symbolic equations (detailed in Appendix~\ref{app::symbolic-regression}) displayed high variance in performance. They surpass expert heuristics on some ideals while failing on others.

Rather than a limitation of the training process, we interpret this result as empirical evidence of the intrinsic complexity of the problem. The optimal ordering is determined by the Gr\"obner fan, a complex structure where ``wall-crossing'' results in discrete jumps in computational cost. The inability of simple interpretable models to generalize suggests that the RL agents are successfully encoding high-dimensional, nonlinear geometric dependencies that defy compression into simple symbolic rules or shallow trees. This validates the necessity of deep reinforcement learning for this task, as the agent navigates a landscape too subtle for standard regression techniques.

\subsection{The ``Sparse Dependence" Heuristic} Despite the complexity of agent strategies, the distillation process revealed a structural insight. Both the decision trees (visualized in Appendix~\ref{app::decision-tree-pictures}) and symbolic expressions tended to rely on a small, specific subset of exponent vectors to make decisions. This indicates that the optimal ordering is often sensitive to a ``critical set" of monomials rather than the entire support. This observation suggests a practical general heuristic for future work: performing careful subsampling of the support set and using this smaller group to decide on a monomial order. 

In summary, while the agent's strategy is irreducible to a simple formula, the distillation confirms that the agent exploits sparse dependencies within the polynomial support to navigate the complex geometry of the problem.

\section{Discussion and Outlook}
We have shown that reinforcement learning provides a flexible and effective framework for optimizing Gr\"obner basis computations, challenging the long-standing reliance on static defaults such as GrevLex. By formulating monomial ordering selection as a learned policy, our agents consistently outperform expert intuition across a range of polynomial systems arising in computer vision and systems biology reducing the cost metric up to 70\%. Our work is the first result that uses machine learning (more specifically RL) on this widely applicable and important problem from computational algebra of selecting monomial orderings. We summarize our basic observations and promising future directions for improvement below.

\textbf{Geometry of Learning.}
Our distillation experiments reveal a notable dichotomy. While reinforcement learning agents are able to navigate the optimization landscape effectively, the strategies they learn resist compression into simple symbolic rules or low-complexity decision models. This behavior suggests that the agents exploit the intricate, high-dimensional structure of the Gr\"obner fan, where effective decisions depend on subtle and nonlinear interactions among features of the polynomial support set.

These observations naturally point beyond the scope of the present study. In particular, they motivate future directions aimed at obtaining a generalizable model across families of problems and to design architectures that are interpretable by design for this particular problem. 

\textbf{Generalizable Models via Graph Neural Networks.}
The agents studied in this work are specialized to fixed polynomial support structures. A natural next step is to pursue more general-purpose neural Gr\"obner optimizers that can transfer across different supports. One promising direction is the use of Graph Neural Networks (GNNs), which can represent polynomial support sets as permutation-invariant graphs encoding relationships among monomials. Such architectures may enable learning policies that generalize beyond individual problem instances, reducing or eliminating the need for per-support training while preserving sensitivity to geometric structure.

\textbf{Interpretable-by-Design Architectures.}
To narrow the gap between agent performance and human interpretability, future work should also explore interpretable-by-design learning architectures. Rather than producing unrestricted weight vectors, agents could be constrained to select from a structured vocabulary of parametrized geometric predicates. By forcing decisions to be expressed in terms of discrete geometric concepts, such models may yield strategies that are both effective and mathematically transparent, offering deeper insight into the geometry underlying efficient Gr\"obner basis computation.

\section*{Acknowledgments}
We thank Tim Duff, Elias Tsigaridas, and Sameer Agarwal for helpful discussions. This work was carried out at the NSF REU Site in Mathematical Analysis and Applications at the University of Michigan--Dearborn. A.E. was partially supported by NSF-CCF-2414160, and all authors were partially supported by the National Science Foundation DMS-2243808.

\bibliographystyle{plain}
\bibliography{rlgrobner}

\newpage
\appendix

\section{Experimental Details} \label{app::experimental-details}

\begin{table}[ht]
  \centering
  \caption{Hyper-parameter choices for the TD3 algorithm and PER buffer.}
  \label{tab:td3-params-full}
    \begin{tabular}{l l l}
      \toprule
      \textbf{Algorithm} & \textbf{Hyper-parameter} & \textbf{Value} \\
      \midrule
      TD3 & Training Episodes & 10,000 \\
          & \(\gamma\) (Reward Discount Factor) & 0.99 \\
          & \(\tau\) (Target Network Update Rate) & 0.05 \\
          & Initial Actor Learning Rate & $10^{-4}$ \\
          & Minimum Actor Learning Rate & $10^{-5}$ \\
          & Initial Critic Learning Rate & $10^{-4}$ \\
          & Minimum Critic Learning Rate & $10^{-6}$ \\
          & $\sigma$ (Exploration Noise Std.) & $0.002$ \\
          & \(d\) (Policy Update Delay) & 100 \\
      \midrule
      PER & Replay Buffer Capacity & 1,000,000 \\
                          & Sampling Batch Size & 100 \\
                          & \(\alpha\) (Prioritization Exponent) & 0.6 \\
                          & \(\beta\) (Importance-Sampling Exponent) & 0.4 \\
                          & \(\eta\) (Beta Annealing Step Size) & \(10^{-4}\) \\
                          & \(\epsilon\) (Priority Offset) & 0.01 \\
      \bottomrule
    \end{tabular}
\end{table}

\textbf{Setup and compute.} We run all experiments with five random seeds per problem instance. The agents were all trained on Google Cloud Compute Engine instances. For the relative pose with unknown focal length problem, three-view triangulation, and \(n\)-site phosphorylation (\(n=14\)) we used a \textit{c4-highcpu-8} instance (8 vCPUs, 16 GB Memory). For the Wnt shuttle model system we used a \textit{c4-highmem-8} instance (8 vCPUs, 62 GB Memory) due to the size of the polynomial support set. We utilized the Twin-Delayed Deep Deterministic Policy Gradient (TD3) algorithm alongside a prioritized experience replay (PER) buffer for the RL agent \cite{fujimotoTD3, schaul2015PER}. 

\textbf{Training procedure.} During training, at each timestep, the actor outputs a real-valued weight vector $w \in \mathbb{R}^n$ (via a softmax), to which we add Gaussian exploration noise; the resulting vector is then discretized into integer weights by scaling by $10^3$, rounding to the nearest integer, and clamping each entry to be at least $1$:
\[
\tilde{w} = \max\!\left(\operatorname{round}\!\left(10^3 \cdot w\right),\, 1\right),\]
where the $\max(\cdot,1)$ is applied element-wise. We then pair $\tilde{w}$ with the corresponding variables to construct a weighted monomial ordering. Using this order, we run Gr\"obner basis computation and obtain a reward given by a Monte Carlo approximation of the reward using sample instances. The agent then updates the weight vector and repeats. An episode consists of a fixed number of these updates and terminates after \(L=25\) steps, and we train for \(N=10{,}000\) episodes. Rewards are averaged over a batch of \(B=10\) ideals per step, where each ideal is instantiated from a fixed polynomial support set and uses uniform random coefficients from the finite field \(\mathbb{F}_{32003} = \mathbb{Z}/32003\mathbb{Z}\). 

\textbf{State and action representation.} The environment state $s_t \in \mathbb{R}^n$ is the current continuous variable weight vector. At the start of each episode, we initialize $s_0$ by sampling $\epsilon_i = 1 + u_i$ with $u_i \sim \mathrm{Unif}(0,1)$ independently and setting $s_0 = \epsilon / \|\epsilon\|_1$. The actor observes a fixed encoding of the polynomial support (the normalized monomial-exponent matrix for the base set) with $s_t$ appended as an additional feature and flattened; the action $a_t$ is a proposed next weight vector (actor output plus exploration noise), and the next state is set to $s_{t+1}=a_t$.

\textbf{Optimization schedule.} Both actor and critic learning rates are linearly decayed from their initial values to their minimum values over the first $90\%$ of training episodes, and are held constant at the minimum for the remaining $10\%$ of episodes. These initial and minimum learning rates are recorded in Table \ref{tab:td3-params-full}.  

\textbf{Evaluation protocol.} We evaluate the trained policy without further learning. We first draw a \emph{calibration batch} of $B=100$ randomly generated ideals. For each ideal, we run the actor in the environment for 25 steps (which matches training) and record the final predicted weight vector, yielding a set of candidate orders. We score each of these candidates by applying it to the entire calibration batch and measuring the mean reward which involves computing Gr\"obner bases for the batch under that fixed order. We select the single best-performing order on this calibration batch. After selecting the final order, we generate a fresh test set of 100,000 new ideals \emph{per seed} and report performance of the selected order on this set. The calibration and test ideals are generated independently using separate RNG seeds. Given the large coefficient field and the number of independently sampled coefficients per ideal, the collision probability is negligible.

\textbf{Network architecture.} We use an actor $\pi_\phi$ and two critics $Q_{\theta_1}, Q_{\theta_2}$ with corresponding target networks as in TD3. Both actor and critics are feed-forward neural networks with three linear hidden layers of width 512 and ReLU activation functions (Table~\ref{tab:nn-arch}). The actor maps the flattened observation vector (support-set features with the current weight state appended) to a length-\(n\) weight vector and applies a softmax. During training, we add Gaussian exploration noise to the actor output prior to discretization. Each critic takes as input the concatenation of the flattened observation and the action vector and outputs a scalar \(Q\)-value.

\begin{table}[tb]
  \centering
  \caption{Neural network architectures used in TD3. We denote the flattened observation dimension by $d_o$.}
  \label{tab:nn-arch}
  \begin{tabularx}{\linewidth}{l X}
    \toprule
    \textbf{Network} & \textbf{Architecture} \\
    \midrule
    Actor $\pi_\phi(o)$ &
    Fully Connected: $d_o \rightarrow 512 \rightarrow 512 \rightarrow 512 \rightarrow n$, ReLU activations, Softmax output. \\

    Critic $Q_{\theta}(o,a)$ (each of two) &
    Fully Connected: $(d_o + n) \rightarrow 512 \rightarrow 512 \rightarrow 512 \rightarrow 1$, ReLU activations. \\
    \bottomrule
  \end{tabularx}
\end{table}

\newpage 
\textbf{Initialization and reproducibility.} Actor and critic networks were implemented in \texttt{Flux.jl} \cite{innes2018Flux}. We used Flux's default parameter initialization (\texttt{glorot\_uniform} for linear layers). Target networks were initialized by copying the online networks prior to soft updates in TD3. We fix random seeds for coefficient sampling, dataset generation, and network initialization/training. Gr\"obner basis computations are performed using \texttt{Groebner.jl}, which may exhibit run-to-run nondeterminism despite seeding.

\newpage
\section{Additional Results} \label{app::additional-results}
In this section we present more detailed results from our experiments involving important ideals arising in computer vision and systems biology. 

\subsection{Relative Pose with Unknown Focal Length} 
\label{app::relative-pose-additional-results}

\begin{figure}[ht]
  \centering
\includegraphics[width=0.5\linewidth]{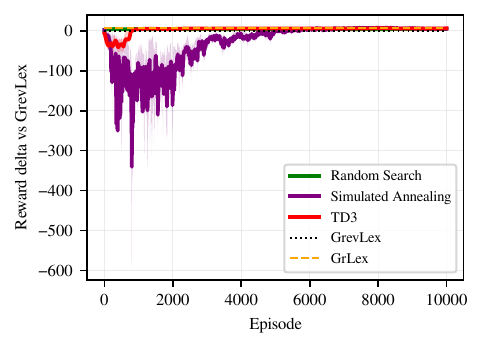}
  \captionsetup{width=0.5\linewidth}
  \caption{Training curves for the relative pose with unknown focal length system. The TD3 agent (red) converges to a stable policy much faster than Simulated Annealing (purple), which requires extensive exploration. Random Search (green), GrevLex, and GrLex baselines are also shown. Interquartile ranges (IQR) across seeds are shaded.}
  \label{fig:relative-pose-training-plot}
\end{figure}

\begin{table}[ht]
\centering
\small
\captionsetup{width=0.85\linewidth}
\caption{Relative Pose: reward robustness across seeds. Entries are median [IQR] across 5 seeds ($10^5$ test ideals per seed). For each seed we compute win/tie/loss rates and the per-seed mean percent change conditional on wins (improvement) or losses (degradation).}
\label{tab:relpose-reward-iqr}
\begin{tabular}{lcc}
\toprule
\textbf{Metric} & \textbf{Agent vs GrevLex} & \textbf{Agent vs GrLex} \\
\midrule
Win rate (\%)  & 98.546 [98.134, 98.551] & 30.993 [28.462, 38.759] \\
Tie rate (\%)  & 0.000  [0.000, 0.000]   & 0.000  [0.000, 37.664] \\
Loss rate (\%) & 1.454  [1.449, 1.866]   & 60.574 [31.343, 61.241] \\
Per-seed mean improvement on wins (\%) & 19.292 [19.280, 19.728] & 8.191 [7.531, 8.428] \\
Per-seed mean degradation on losses (\%) & -1.411 [-1.444, -0.957] & -7.366 [-10.028, -7.284] \\
\bottomrule
\end{tabular}
\end{table}

\begin{table}[ht]
\centering
\captionsetup{width=0.87\linewidth}
\caption{Final discretized agent weight vectors for the Relative Pose with Unknown Focal Length system across five random seeds. We report the integer weight vector $\tilde w=\max(\mathrm{round}(10^3 \cdot w),1)$ used to define the monomial order and the implied variable priority order. Bold indicates the seed with the largest estimated mean reward improvement vs.\ GrevLex, computed as $\widehat{\Delta}=(\mathrm{win}/100)\cdot(\mathrm{improve\_win\_only})+(\mathrm{loss}/100)\cdot(\mathrm{degrade\_loss\_only})$ (ties contribute $0$).}
\label{tab:relative-pose-final-weights}
\resizebox{0.5\linewidth}{!}{%
\begin{tabular}{c l c}
\toprule
Seed & Final discretized weights $\tilde w$ & Implied variable order \\
\midrule
0 & (373, 305, 323) & $x_{1} > x_{3} > x_{2}$ \\
1 & (322, 297, 382) & $x_{3} > x_{1} > x_{2}$ \\
2 & (308, 313, 379) & $x_{3} > x_{2} > x_{1}$ \\
\textbf{3} & \textbf{(326, 352, 322)} & \boldmath $x_{2} > x_{1} > x_{3}$ \unboldmath \\
4 & (318, 373, 309) & $x_{2} > x_{1} > x_{3}$ \\
\bottomrule
\end{tabular}%
}
\end{table}

\FloatBarrier

\subsection{Three-View Triangulation}
\label{app::triangulation-additional-results}

\begin{figure}[H]
  \centering
  \includegraphics[width=0.5\linewidth]{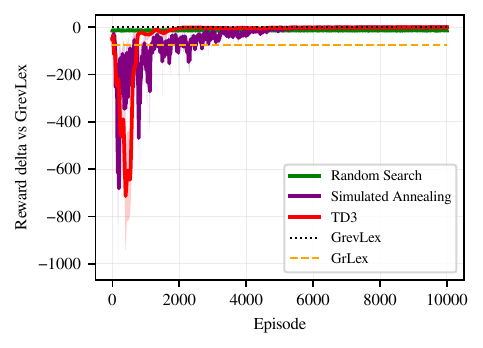}
  \captionsetup{width=0.5\linewidth}
  \caption{Training curves for the three-view triangulation system. The TD3 agent (red) converges to a stable policy much faster than Simulated Annealing (purple), which requires extensive exploration. Random Search (green), GrevLex, and GrLex baselines are also shown. Interquartile ranges (IQR) across seeds are shaded.}
  \label{fig:triangulation-training-plot}
\end{figure}

\begin{table}[H]
\centering
\small
\captionsetup{width=0.85\linewidth}
\caption{Triangulation: reward robustness across seeds. Entries are median [IQR] across 5 seeds ($10^5$ test ideals per seed). For each seed we compute win/tie/loss rates and the per-seed mean percent change conditional on wins (improvement) or losses (degradation).}
\label{tab:triangulation-reward-iqr}
\begin{tabular}{lcc}
\toprule
\textbf{Metric} & \textbf{Agent vs GrevLex} & \textbf{Agent vs GrLex} \\
\midrule
Win rate (\%)  & 8.565 [8.486, 9.343]     & 100.000 [100.000, 100.000] \\
Tie rate (\%)  & 82.822 [81.176, 83.136]  & 0.000 [0.000, 0.000] \\
Loss rate (\%) & 8.378 [8.322, 8.613]     & 0.000 [0.000, 0.000] \\
Per-seed mean improvement on wins (\%) & 0.715 [0.715, 0.716] & 37.509 [37.508, 37.510] \\
Per-seed mean degradation on losses (\%) & -0.724 [-0.724, -0.721] & -- \\
\bottomrule
\end{tabular}
\end{table}

\begin{table}[H]
\centering
\captionsetup{width=0.85\linewidth}
\caption{Final discretized agent weight vectors for the Three-View Triangulation system across five random seeds. We report the integer weight vector $\tilde w=\max(\mathrm{round}(10^3 w),1)$ used to define the monomial order and the implied variable priority order. Bold indicates the seed with the largest estimated mean reward improvement vs.\ GrevLex, computed as $\widehat{\Delta}=(\mathrm{win}/100)\cdot(\mathrm{improve\_win\_only})+(\mathrm{loss}/100)\cdot(\mathrm{degrade\_loss\_only})$ (ties contribute $0$).}
\label{tab:triangulation-final-weights}
\resizebox{0.5\linewidth}{!}{%
\begin{tabular}{c l c}
\toprule
Seed & Final discretized weights $\tilde w$ & Implied variable order \\
\midrule
0 & (230, 401, 369) & $x_{2} > x_{3} > x_{1}$ \\
1 & (311, 348, 340) & $x_{2} > x_{3} > x_{1}$ \\
2 & (371, 372, 257) & $x_{2} > x_{1} > x_{3}$ \\
3 & (264, 365, 371) & $x_{3} > x_{2} > x_{1}$ \\
\textbf{4} & \textbf{(355, 305, 340)} & \boldmath $x_{1} > x_{3} > x_{2}$ \unboldmath \\
\bottomrule
\end{tabular}%
}
\end{table}

\FloatBarrier

\subsection{\(n\)-Site Phosphorylation System}
\label{app::n-site-additional-results}

\begin{figure}[H]
  \centering
  \includegraphics[width=0.5\linewidth]{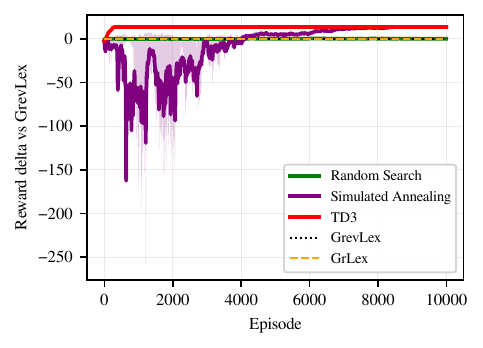}
  \captionsetup{width=0.5\linewidth}
  \caption{Training curves for the \(n\)-site phosphorylation system. The TD3 agent (red) converges to a stable policy much faster than Simulated Annealing (purple), which requires extensive exploration. Random Search (green), GrevLex, and GrLex baselines are also shown. Interquartile ranges (IQR) across seeds are shaded.}
  \label{fig:n-site-training-plot}
\end{figure}

\begin{table}[H]
\centering
\small
\captionsetup{width=0.85\linewidth}
\caption{N-site phosphorylation: reward robustness across seeds. Entries are median [IQR] across 5 seeds ($10^5$ test ideals per seed). For each seed we compute win/tie/loss rates and the per-seed mean percent change conditional on wins (improvement) or losses (degradation).}
\label{tab:nsite-reward-iqr}
\begin{tabular}{lcc}
\toprule
\textbf{Metric} & \textbf{Agent vs GrevLex} & \textbf{Agent vs GrLex} \\
\midrule
Win rate (\%)  & 100.000 [100.000, 100.000] & 100.000 [100.000, 100.000] \\
Tie rate (\%)  & 0.000 [0.000, 0.000]       & 0.000 [0.000, 0.000] \\
Loss rate (\%) & 0.000 [0.000, 0.000]       & 0.000 [0.000, 0.000] \\
Per-seed mean improvement on wins (\%) & 70.770 [70.770, 70.771] & 70.770 [70.769, 70.771] \\
Per-seed mean degradation on losses (\%) & -- & -- \\
\bottomrule
\end{tabular}
\end{table}

\begin{table}[ht]
\centering
\captionsetup{width=0.85\linewidth}
\caption{Final discretized agent weight vectors for the \(n\)-Site Phosphorylation system across five random seeds. We report the integer weight vector $\tilde w=\max(\mathrm{round}(10^3 w),1)$ used to define the monomial order and the implied variable priority order. Bold indicates the seed with the largest estimated mean reward improvement vs.\ GrevLex, computed as $\widehat{\Delta}=(\mathrm{win}/100)\cdot(\mathrm{improve\_win\_only})+(\mathrm{loss}/100)\cdot(\mathrm{degrade\_loss\_only})$ (ties contribute $0$).}
\label{tab:nsite-final-weights}
\resizebox{0.5\linewidth}{!}{%
\begin{tabular}{c l c}
\toprule
Seed & Final discretized weights $\tilde w$ & Implied variable order \\
\midrule
0 & (1000, 1) & $x_{1} > x_{2}$ \\
\textbf{1} & \textbf{(1000, 1)} & \boldmath $x_{1} > x_{2}$ \unboldmath \\
2 & (1000, 1) & $x_{1} > x_{2}$ \\
3 & (955, 45) & $x_{1} > x_{2}$ \\
4 & (983, 17) & $x_{1} > x_{2}$ \\
\bottomrule
\end{tabular}%
}
\end{table}

\FloatBarrier

\subsection{Wnt Shuttle Model}
\label{app::wnt-additional-results}

\begin{figure}[H]
  \centering
  \includegraphics[width=0.5\linewidth]{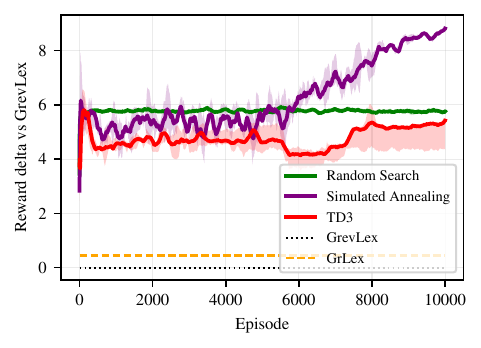}
  \captionsetup{width=0.5\linewidth}
  \caption{Training curves for the Wnt shuttle model system. The TD3 agent (red) converges to a stable policy much faster than Simulated Annealing (purple), which requires extensive exploration. Random Search (green), GrevLex, and GrLex baselines are also shown. Interquartile ranges (IQR) across seeds are shaded.}
  \label{fig:wnt-training-plot}
\end{figure}

\begin{table}[H]
\centering
\small
\captionsetup{width=0.85\linewidth}
\caption{WNT: reward robustness across seeds. Entries are median [IQR] across 5 seeds ($10^5$ test ideals per seed). For each seed we compute win/tie/loss rates and the per-seed mean percent change conditional on wins (improvement) or losses (degradation).}
\label{tab:wnt-reward-iqr}
\begin{tabular}{lcc}
\toprule
\textbf{Metric} & \textbf{Agent vs GrevLex} & \textbf{Agent vs GrLex} \\
\midrule
Win rate (\%)  & 99.850 [99.212, 100.000] & 96.475 [96.252, 100.000] \\
Tie rate (\%)  & 0.000 [0.000, 0.000]     & 0.000 [0.000, 0.000] \\
Loss rate (\%) & 0.150 [0.000, 0.788]     & 3.525 [0.000, 3.748] \\
Per-seed mean improvement on wins (\%) & 39.319 [37.646, 69.656] & 38.010 [35.288, 68.035] \\
Per-seed mean degradation on losses (\%) & -8.274 [-17.562, -7.442] & -3.786 [-13.418, -3.079] \\
\bottomrule
\end{tabular}
\end{table}

\begin{table}[tb]
\centering
\small
\setlength{\tabcolsep}{4pt}
\caption{Final discretized agent weight vectors for the Wnt Shuttle Model system across five random seeds. We report the integer weight vector $\tilde w=\max(\mathrm{round}(10^3 w),1)$ (only non-unit entries shown for readability; omitted entries equal $1$) and the implied variable priority order. Bold indicates the seed with the largest estimated mean reward improvement vs.\ GrevLex, computed as $\widehat{\Delta}=(\mathrm{win}/100)\cdot(\mathrm{improve\_win\_only})+(\mathrm{loss}/100)\cdot(\mathrm{degrade\_loss\_only})$ (ties contribute $0$).}
\label{tab:wnt-final-weights}
\begin{tabularx}{\linewidth}{c X X}
\toprule
Seed & Final discretized weights $\tilde w$ (sparse; non-ones) & Implied variable order \\
\midrule
0 & $x_{19}\!:\!1000$ (others $1$) & $x_{19} > \text{others (tied)}$ \\
1 & $x_{17}\!:\!1000$ (others $1$) & $x_{17} > \text{others (tied)}$ \\
2 & $x_{2}\!:\!512,\; x_{5}\!:\!434,\; x_{17}\!:\!11,\; x_{3}\!:\!10,\; x_{12}\!:\!9,\; x_{7}\!:\!6,\; x_{18}\!:\!6,\; x_{14}\!:\!5,\; x_{10}\!:\!2$ (others $1$)
  & $x_{2} > x_{5} > x_{17} > x_{3} > x_{12} > (x_{7}=x_{18}) > x_{14} > x_{10} > \text{others}$ \\
3 & $x_{2}\!:\!1000$ (others $1$) & $x_{2} > \text{others (tied)}$ \\
\textbf{4} & \boldmath $x_{16}\!:\!494,\; x_{17}\!:\!394,\; x_{13}\!:\!79,\; x_{5}\!:\!21,\; x_{2}\!:\!5,\; x_{19}\!:\!4,\; x_{7}\!:\!2$ \unboldmath \textbf{(others $1$)}
  & \boldmath \textbf{$x_{16} > x_{17} > x_{13} > x_{5} > x_{2} > x_{19} > x_{7} > \text{others}$} \unboldmath \\
\bottomrule
\end{tabularx}
\end{table}
\FloatBarrier

\clearpage

\section{Symbolic Regression Equations} \label{app::symbolic-regression}

\subsection{Best Symbolic Regression Equations} \label{app::symbolic-best-equations}

\begin{equation}
 w_1(x)
=
\frac{0.005620111648365623 \, x_{15}}
     {0.41764189247505734 - (x_{42} - x_{21}) x_1}
+ 0.3285121764974553
\end{equation}

\begin{center}
(see Table~\ref{csv1}, Complexity 13)
\end{center}

\begin{multline}
w_2(x)
=
-0.002235617301013779
\Bigg[
\frac{x_{40} - (x_8 - x_{43})}
     {x_{29} x_{33} + 0.23181594905978833}
\\
- \Big(
x_{33} x_9
+ x_{29}
  \frac{-2.6092572238752862 \, x_{31}}
       {x_{24} + 0.8938154468830032}
- (x_{51} x_{46} + x_{47} x_{37})
\Big)
\Bigg]
+ 0.3591951413920924
\end{multline}

\begin{center}
(see Table~\ref{csv2}, Complexity 37)
\end{center}

\begin{multline}
w_3(x)
=
0.0025153236949469683
\Bigg[
\big(x_{35} - (x_{45} - x_{18})\big)
\Big(
x_{43}
- 0.1117044718447674 \, x_{24} x_{15} x_{10}
\Big)+ (x_{46} - x_{22}) x_{47} \\
+ \Big(
x_{45}
- \big(
(x_{51} - (x_{46} - 2.472476686808669) - x_{21})
\, 0.4030994964706104
\big)
- x_{15}
\Big)^2
\Bigg]
+ 0.3125372513473696
\end{multline}

\begin{center}
(see Table~\ref{csv3}, Complexity 40)
\end{center}

\subsection{Symbolic Regression Runner-ups} 
\label{app::symbolic-halloffame}
The following tables report the Hall of Fame expressions obtained through \texttt{SymbolicRegression.jl}. For each target, expressions are shown from the Pareto frontier, where complexity is measured by node count with each operator, variable, and constant contributing unit cost, and loss corresponds to mean squared error (L2DistLoss). Expressions are ordered by increasing complexity, and for each complexity level the expression with the lowest mean squared error is reported.

\begin{longtable}{r l p{0.68\linewidth}}
\caption{Best equations at each complexity for \(x_1\) weight.} \label{csv1} \\
\hline
\textbf{Complexity} & \textbf{Loss} & \textbf{Equation} \\ \hline
\endfirsthead
1  & 0.00084203 & \(0.335353908454586\) \\

5  & 0.00080188 & \(-0.0046158769662659116\,x_{33} + 0.34548194929346504\) \\

6  & 0.00080159 &
\(\big(-0.0022266810459186485\,x_{33} + 0.8085746701887174\big)^5\) \\

7  & 0.00076023 &
\(0.007101505925913343 / (0.5008673141532315 - x_1) + 0.33771927404824137\) \\

8  & 0.00075939 &
\(0.006882074640602726 / (0.39837595753768223 - x_1^4) + 0.33472540695934044\) \\

9  & 0.00070293 &
\(x_{15}\,0.00031468880927016735 / (0.02391101342623325 - x_1) + 0.33117132434738383\) \\

10 & 0.00065630 &
\(-0.007372465610914996 / (x_3 - (x_1 + 0.5591087934678365)^3) + 0.3340533898078061\) \\

12 & 0.00062627 &
\(x_{15}\,0.0015838879526246991 / (x_3 - (x_1 - 0.4366839685254661)^3) + 0.3330674836798489\) \\
\rowcolor{green!15}
13\label{thirteen} & 0.00056414 &
\(0.005620111648365623\,x_{15} / (0.41764189247505734 - (x_{42}-x_{21})x_1)
+ 0.3285121764974553\) \\

14 & 0.00055918 &
\(0.0017032708532298386\,x_{15}^2 / (0.42069267319326054 - (x_{42}-x_{21})x_1)
+ 0.3301019930946452\) \\

15 & 0.00054388 &
\(0.00797284069124181\,(x_{15}-0.8129971943825058) \allowbreak /
(0.4487955698570075 - (x_{42}-x_{21})x_1)
+ 0.33110916450723943\) \\

16 & 0.00051155 &
\(-0.001985966203118873\,(x_3 - x_{15}^2) /
(0.405649388473595 - (x_{42}-x_{21})x_1)
+ 0.3316337474494216\) \\

17 & 0.00049757 &
\((x_{15} - 0.45311719799127026\,x_3)\,
(0.007098223286061685 / \allowbreak
(0.40730076571579676 - (x_{42}-x_{21})x_1))
+ 0.3308134115589299\) \\

18 & 0.00049701 &
\(\big(
0.0034905131345707146\,(x_{15}-0.4618026758965021\,x_3) /
(0.4312939940164463 - (x_{42}-x_{21})x_1)
+ 0.8015132923036502
\big)^5\) \\

19 & 0.00048087 &
\(0.006779004173437662\,(x_{15}-0.24535182897116273\,x_{11}x_3) / \allowbreak
(0.4353455212996979 - (x_{42}-x_{21})x_1)
+ 0.3309002658196576\) \\

20 & 0.00047959 &
\(\big(
-0.5752651687819168
- 0.005957292217184054\,(x_{15}-0.25527858194238934\,x_{11}x_3) /
(0.4500364214761212 - (x_{42}-x_{21})x_1)
\big)^2\) \\

21 & 0.00046960 &
\(-0.013717485498834751\,x_{15} /
((x_{32}((x_{25}-0.4069727000080694\,x_{34})x_3))
- (0.8404253408504924 - x_1 x_7))
+ 0.3339296560732781\) \\

22 & 0.00046185 &
\(\big(
-0.5745563514052511
- 0.0054471371796765\,
(x_{15} + 0.3239162142722927(x_{17}-x_{11}x_3)) /
(0.47631273534124813 - (x_{42}-x_{21})x_1)
\big)^2\) \\

23 & 0.00045559 &
\(x_{15}\,
(-0.014214964317353139 /
(x_1 x_7 + x_{32}((x_{25}-0.40063961198015563\,x_{34})x_3)
- 0.9081076188299587)
+ 0.002607236858603949)
+ 0.32883104309502337\) \\

24 & 0.00044210 &
\(\big(
-0.5755104167273538
- 0.006397558424129668\,
(x_{15}+0.3203210787099133((x_{11}-x_{43})-x_{11}x_3)) /
(0.4769463584228517 - (x_{42}-x_{21})x_1)
\big)^2\) \\

26 & 0.00044192 &
\(\big(
(x_{15}+0.3097648286613735((x_{11}-x_{11}x_3)-(x_{43}+0.24942426276230753)))
(0.0065583064541704225 / \allowbreak
(0.47852394857981884 - (x_{42}-x_{21})x_1))
+ 0.575664307607853
\big)^2\) \\

27 & 0.00043258 &
\((-0.014657285921587318 /
(x_7 x_1 + x_{32}((x_2-x_7-0.3960833348371273\,x_{34})(x_3-0.04753016637230385))
- 0.9318601830011407)
+ 0.0027135047221870136)x_{15}
+ 0.3279745155651019\) \\

28 & 0.00042895 &
\(\big(
-0.5743323359043249
- (x_{15}+0.3115162090541634((x_{11}-x_{11}x_3)-((x_{43}+1.8546631396767344)-x_{22})))
(0.006476643266282883 /
(0.4822291194014101 - (x_{42}-x_{21})x_1))
\big)^2\) \\
\end{longtable}

\begin{longtable}{r l p{0.68\linewidth}}
 \caption{Best equations at each complexity for \(x_2\) weight.\label{csv2}}\\
\hline
\textbf{Complexity} & \textbf{Loss} & \textbf{Equation} \\
\hline
\endhead

\hline
\endfoot

1  & 0.00078947 & \(0.333991507716237\) \\

5  & 0.00075544 & \(-0.004199115704102301\,x_{46} + 0.3425935796304692\) \\

6  & 0.00075529 & \(\big(-0.0020277713924658834\,x_{46} + 0.8071924178529383\big)^5\) \\

7  & 0.00072033 & \(-0.0021130167799510383\,x_{47}x_{31} + 0.3403100433298779\) \\

8  & 0.00071989 & \(\big(0.8061166767001957 - 0.0010366330129766433\,x_{31}x_{47}\big)^5\) \\

9  & 0.00066067 & \(0.0010987382178416394\,x_{31}x_{46}x_{47} + 0.34090395786341393\) \\

10 & 0.00065699 &
\(-0.0003371555036407168\,x_{47}\,x_{31}\,x_{46}^2 + 0.34008649358858045\) \\

11 & 0.00062943 &
\(0.0020338646068404674\,x_{31}\,(x_{19}-x_{46}-x_{47}) + 0.3407644743185726\) \\

13 & 0.00060744 &
\(0.0021405969477780082\,(x_{45}+x_{31}(x_{19}-x_{46}-x_{47})) + 0.3363607121050117\) \\

15 & 0.00058262 &
\(0.0025081350655382258\,(x_{45}+x_{31}(x_{9}-x_{46}-x_{47})+x_{33}) + 0.33097200530614007\) \\

17 & 0.00055884 &
\(0.00191878170741938\Big((x_{31}x_{46}-x_{45})/(-0.3060438988542676-x_{24}) - x_{37}x_{47}\Big)
+ 0.3451167234987364\) \\

19 & 0.00051931 &
\(\Big(x_{46}x_{51} - 0.0022148380626643926\big(x_{37}-(x_{33}+x_{31}/(-0.44420884210847517-x_{24}))x_{39}\big)\Big)
+ 0.3490547776658768\) \\

20 & 0.00051481 &
\(\Big(x_{51}x_{46} - 0.0023200756665093906\big(x_{37}-(x_{31}/(-0.781593885421935-x_{24})^3+x_{33})x_{39}\big)\Big)
+ 0.3471331687446784\) \\

21 & 0.00048898 &
\(0.00236733598928659\Big(((x_{31}/(-0.456255437752778-x_{24})+x_{33})-x_{37})x_{39} - (x_{50}+x_{46}x_{51})\Big)
+ 0.3551263876805231\) \\

22 & 0.00048675 &
\(0.002443052189456914\Big(x_{39}\big((x_{33}-x_{37})+x_{31}/(-0.7859361386370655-x_{24})^3\big) - (x_{51}x_{46}+x_{50})\Big)
+ 0.35308082825021847\) \\

23 & 0.00046503 &
\(0.0018248037128027186\Big(((x_{31}/(-0.38099333741954855-x_{24})+x_{33}-x_{50})x_{39}) - (x_{47}x_{37}+x_{51}x_{46})\Big)
+ 0.35378347325409515\) \\

24 & 0.00046297 &
\(\Big(-0.5949846470208265 - 0.001586193251114419\big(x_{39}(x_{33}+x_{31}/(-0.37200420537481116-x_{24})-x_{50})-(x_{51}x_{46}+x_{47}x_{37})\big)\Big)^2\) \\

25 & 0.00044263 &
\(\Big(x_{46}x_{51} - 0.001914639887438814\big((x_{37}-(x_{8}-x_{2}))x_{47} - x_{39}(x_{31}/(-0.382712551548299-x_{24})+x_{33})\big)\Big)
+ 0.3473789455394515\) \\

26 & 0.00044159 &
\(\Big(-0.5893700679989564 + 0.0016580897566988162\big((x_{47}(x_{37}-(x_{8}-x_{2}))+x_{46}x_{51}) - x_{39}(x_{31}/(-0.373203831216098-x_{24})+x_{33})\big)\Big)^2\) \\

27 & 0.00042545 &
\(\Big((x_{47}(x_{37}-(x_{8}-x_{2})) + x_{51}x_{46}) - (x_{33}+(x_{31}/(-0.3321670393695553-x_{24})-x_{12}))x_{39}\Big)(-0.0018145013437147934)
+ 0.35446315781549986\) \\

29 & 0.00041140 &
\(\Big(((x_{37}-(x_{8}-x_{41}))x_{47}) - (x_{23} + (x_{33}+x_{31}/(-0.33046765097746766-x_{24})-x_{12})x_{39}) + x_{46}x_{51}\Big)(-0.0018098846467069424)
+ 0.35019788451166994\) \\

31 & 0.00041134 &
\(\Big(((x_{37}-(x_{8}-x_{41}))x_{47}) - (x_{23} + (x_{33}+x_{31}/(-0.33046765097746766-x_{24})-(x_{12}-0.17452676144906673))x_{39}) + x_{46}x_{51}\Big)(-0.0018098846467069424)
+ 0.35019788451166994\) \\

32 & 0.00040966 &
\(\Big(x_{43}/(x_{29}^3+0.21668132336426366)
- \big((x_{33}x_{9}) + x_{29}((x_{31}/(x_{24}+0.9799775476388732))(-3.1441422538653527)) - (x_{46}x_{51}+x_{24}x_{47})\big)\Big)(-0.002171787760626936)
+ 0.35919383963827073\) \\

33 & 0.00038691 &
\(0.0021205150448707823\Big(
((x_{31}x_{29})/(x_{24}+0.8466651252176755))(-2.614565853580865)
- (x_{51}x_{46}+x_{47}x_{37})
+ x_{9}x_{33}
- x_{40}/(x_{29}x_{33}+0.19545647402663877)
\Big)
+ 0.3594572066076611\) \\

35 & 0.00035425 &
\(0.002244557052296696\Big(
x_{9}x_{33} - (x_{51}x_{46}+x_{47}x_{37})
+ ((x_{29}x_{31})(-2.5146823106334244))/(x_{24}+0.8636999456811458)
- (x_{40}-(x_{8}-x_{43}))/(x_{29}+0.22228327081320715)
\Big)
+ 0.35955925873265626\) \\

36 & 0.00035230 &
\(0.002272116513640928\Big(
((-2.6090149778032843/(x_{24}+0.8973959852574365))x_{29})x_{31}
- (x_{47}x_{37}+x_{46}x_{51})
+ x_{33}x_{9}
- (x_{40}-(x_{8}-x_{43}))/(x_{29}^2+0.2264743324062024)
\Big)
+ 0.3592286936836383\) \\
\rowcolor{green!15}
37 & 0.00033215 &
\(-0.002235617301013779\Big(
(x_{40}-(x_{8}-x_{43}))/(x_{29}x_{33}+0.23181594905978833)
- \big(x_{33}x_{9} + (x_{29}((x_{31}(-2.6092572238752862))/(x_{24}+0.8938154468830032)) - (x_{51}x_{46}+x_{47}x_{37}))\big)
\Big)
+ 0.3591951413920924\) \\

38 & 0.00033147 &
\(0.002244425401757337\Big(
(x_{29}x_{31}(-2.6093339340107944))/(x_{24}+0.8945511162213898)
+ \big(x_{9}x_{33}-(x_{46}x_{51}+x_{47}x_{37})\big)
- (x_{40}-(x_{8}-x_{43}))/(x_{33}x_{29}^2+0.2339784392837013)
\Big)
+ 0.35891790936256646\) \\

40 & 0.00033098 &
\(\Big(
(x_{51}x_{46}+x_{37}x_{47})
- \big((x_{45}-x_{53})
- (x_{40}-(x_{8}-x_{43}))/(x_{29}^4+0.2417138822628581)
+ x_{33}x_{9}
+ (x_{29}x_{31}(-2.7453391278876245))/(x_{24}+0.9560107774969062)\big)
\Big)(-0.00230981887182462)
+ 0.3588645590624389\) \\

41 & 0.00032537 &
\(0.0021996038348656905\Big(
x_{33}(x_{9}+0.9992791933740277)
+ x_{31}\big(x_{29}(-2.829653528461395/(x_{24}+0.9326840103556284))\big)
- (x_{37}x_{47}+x_{51}x_{46})
+ x_{45}
- (x_{40}-(x_{8}-x_{43}))/(x_{29}+0.22762105258676915)
- x_{53}
\Big)
+ 0.3545816332666742\) \\

42 & 0.00032388 &
\(0.002233692909954876\Big(
(x_{9}+0.9192291261504111)x_{33}
+ \big(x_{29}(x_{31}(-2.8589974979019375/(x_{24}+0.9481617496027702))) - (x_{37}x_{47}+x_{51}x_{46}) + x_{45} - x_{53}\big)
- (x_{40}-(x_{8}-x_{43}))/(x_{29}^2+0.23352143425750302)
\Big)
+ 0.35450471133492284\) \\

43 & 0.00031733 &
\(0.002338160881345756\Big(
((x_{31}(-2.8152793987412856)-x_{43}(-0.7773759055994179))x_{29})/(x_{24}+0.8522483091384043)
+ (x_{45}-x_{53})
- (x_{47}x_{37}+x_{46}x_{51})
+ x_{33}x_{9}
- (x_{40}-(x_{8}-x_{43}))/(x_{29}+0.2500061705458608)
\Big)
+ 0.35790957885390356\) \\

45 & 0.00030834 &
\(0.0023014495193051394\Big(
x_{45} - (x_{40}-(x_{8}-x_{43}))/(x_{29}+0.2447808603451855)
+ (x_{9}+0.8456580509874632)x_{33} - (x_{51}x_{46}+x_{47}x_{37})
+ ((x_{31}(-2.827475727728687)-x_{43}(-0.7779616422592833))x_{29})/(x_{24}+0.819637925080654)
- x_{53}
\Big)
+ 0.3537012271375328\) \\

47 & 0.00030677 &
\(0.0023273398597252405\Big(
x_{45} + (x_{9}+0.7608004910166554)x_{33}
- (x_{47}x_{37}+x_{51}x_{46})
- x_{53}
+ ((x_{31}(-2.8319135217202875)-x_{43}(-0.7538195202526867))x_{29})/(x_{24}+0.82042160941259)
- ((x_{40}+0.5594128253067995)-(x_{8}-x_{43}))/(x_{29}+0.27076459337861525)
\Big)
+ 0.3554408433301304\) \\

49 & 0.00030674 &
\(0.9115388333286707\Big(
0.002544728786040382\Big(
x_{45} + (x_{9}+0.7653057555117491)x_{33} - (x_{51}x_{46}+x_{37}x_{47})
- x_{53}
+ (x_{29}(x_{31}(-2.834934086133134)-x_{43}(-0.7514412054558395)))/(x_{24}+0.821887430914458)
- (x_{40}+(0.5595340556515455-(x_{8}-x_{43})))/(x_{29}+0.2715583381850336)
\Big)
+ 0.3897378324018952
\Big)\) \\

\end{longtable}

\begin{longtable}{r l p{0.68\linewidth}}
\caption{Best equations at each complexity for \(x_3\) weight.}
\label{csv3} \\

\hline
\textbf{Complexity} & \textbf{Loss} & \textbf{Equation} \\
\hline
\endhead

\hline
\endfoot

1  & 0.00087608 & \(0.33065458382200147\) \\

5  & 0.00082794 & \(0.0049947168548858814\,x_{46} + 0.320422688130653\) \\

6  & 0.00082254 & \(8.042901951560854\times 10^{-5}\,x_{43}^4 + 0.3254954135905829\) \\

7  & 0.00078902 & \(0.004545560610712497\,(x_{46}+x_{43}) + 0.31256060469366886\) \\

8  & 0.00075318 & \(0.3366298173974247 - 1.0092803392911087\times 10^{-5}\,x_{15}\,x_{21}^5\) \\

9  & 0.00073919 & \(0.33131528802674354 - 0.0021952429973046412\,x_{15}\,(x_{21}-x_{51})\) \\

10 & 0.00069402 & \(0.33157528202583064 + 4.5834661596589784\times 10^{-5}\,x_{15}^4\,(x_{51}-x_{21})\) \\

11 & 0.00065058 & \(0.3325734057882249 - 0.004033442088750746\,(x_{51}-x_{21})(x_{1}-x_{15})\) \\

12 & 0.00063675 & \(8.56259399681715\times 10^{-5}\,x_{5}\,(x_{51}-x_{21})\,x_{15}^3 + 0.3324602104413529\) \\

13 & 0.00059392 & \(0.3312978884974693 + 0.002175888564310824\,x_{5}\,(x_{51}-x_{21})\,(x_{15}-1.6545830265253603)\) \\

14 & 0.00058819 & \(0.3321442926473853 + 0.0010229333922801703\,(x_{15}^2 - x_{1}x_{45})(x_{51}-x_{21})\) \\

15 & 0.00055270 &
\(0.33833886120375295 + 0.002129155628770091\,x_{5}\big((x_{51}-x_{21})(x_{15}-1.8194694713075403)-x_{11}\big)\) \\

16 & 0.00055264 &
\(\big(0.5815316989581872 + 0.0018580403870196221\,x_{5}\big((x_{51}-x_{21})(x_{15}-1.8063902347770566)-x_{11}\big)\big)^2\) \\

17 & 0.00049840 &
\(0.33044428516462354 + 0.0022218664448306253\,x_{5}\big(((x_{15}-1.9345985168978028)(x_{51}-x_{21})+x_{43})-x_{11}\big)\) \\

19 & 0.00047904 &
\(0.3355111863721219 - 0.0023188669021414637\big(x_{5}(x_{11}-(x_{43}+(x_{51}-x_{21})(x_{15}-1.9273556047791913)))+x_{39}\big)\) \\

21 & 0.00046841 &
\(0.33842817649545387 - 0.0022827681332977198\big(x_{5}(x_{11}-(x_{43}+(x_{51}-x_{21})(x_{15}-1.9897227444584367)))+x_{39}x_{8}\big)\) \\

23 & 0.00046485 &
\(-0.0022684016080321712\Big(x_{22}-x_{5}\big((x_{51}-x_{21})(x_{15}-1.8943375090532004)+x_{43}-0.3957861122866493\,x_{12}x_{11}\big)\Big)
+ 0.3327055940376386\) \\

25 & 0.00045836 &
\(0.3331880635541003 + 0.002229994470279145\,x_{5}\Big((x_{15}-1.9168858125859567)(x_{51}-x_{21})
-0.36987256794841716\,x_{12}x_{11}
+ x_{43} - x_{44}/1.7948234686692892\Big)\) \\

26 & 0.00045676 &
\(0.3352609855215257 + 0.000968590126591488\Big((x_{12}-0.6265352014847679)\,x_{26}\,(x_{46}-x_{11}-x_{8})
+ (x_{51}-x_{21})(x_{15}^2-x_{45}x_{1})\Big)\) \\

27 & 0.00044707 &
\(0.33148505384565896 + 0.0025266111034353417\,x_{5}\Big(-0.3125198732383364\,x_{11}x_{12}
+ (x_{51}-x_{21})(x_{15}-1.9064262470349407)
+ x_{43} - x_{26}/(x_{9}+1.1262315507580507)\Big)\) \\

28 & 0.00044220 &
\(0.002816834926431939\Big((x_{35}-(x_{45}-x_{18}))\big(x_{43}-0.26704001906409974\,x_{15}x_{10}\big)
+ x_{47}(x_{46}-x_{22}) + (x_{45}-x_{15})^2\Big)
+ 0.3136529125025496\) \\

29 & 0.00043477 &
\(0.33148250482799 + 0.002432353906894396\,x_{5}\Big(-0.30819736508643114\,x_{12}x_{11}
+ x_{43} + (x_{51}-x_{21})(x_{15}-1.9237359137999686)
- x_{26}/(x_{9}x_{46}+1.1512804582845528)\Big)\) \\

30 & 0.00038809 &
\(0.002743153299578441\Big((x_{35}-(x_{45}-x_{18}))\big(x_{43}-0.10659040733797036\,x_{24}x_{10}x_{15}\big)
+ x_{47}(x_{46}-x_{22}) + (x_{15}-x_{45})^2\Big)
+ 0.3138307221906743\) \\

32 & 0.00037873 &
\(0.0028183583126295706\Big((x_{35}-(x_{45}-x_{18}))\big(x_{43}-0.11116423563988116\,x_{24}x_{15}x_{10}\big)
+ (x_{46}-x_{22})x_{47} + x_{24} + (x_{15}-x_{45})^2\Big)
+ 0.30741446558876223\) \\

34 & 0.00036038 &
\(0.0023427157748366328\Big(\big(((x_{45}-(x_{51}-x_{21}))0.7292014270107886-x_{15})^2\big)
+ x_{45}(x_{46}-x_{22})
+ \big((x_{35}-(x_{45}-x_{18}))(x_{43}-0.28395945551346546\,x_{10}x_{15})\big)\Big)
+ 0.31289431057379996\) \\

36 & 0.00033513 &
\(0.0025656042317600243\Big(x_{47}(x_{46}-x_{22})
+ \big((x_{15}-(x_{45}-0.3808200991158191(x_{51}-x_{21})))^2
+ (x_{35}-(x_{45}-x_{18}))\big(x_{43}-0.10286795355494745\,x_{15}x_{10}x_{24}\big)\big)\Big)
+ 0.3123325801690137\) \\

38 & 0.00032179 &
\(0.0023279484917965547\Big(x_{10}
+ \big((0.7053528913176984(x_{45}-(x_{51}-x_{21}))-x_{15})^2\big)
+ x_{47}(x_{46}-x_{22})
+ (x_{35}-(x_{45}-x_{18}))\big(x_{43}-0.12042537845498498\,x_{15}x_{24}x_{10}\big)\Big)
+ 0.30903266715539784\) \\
\rowcolor{green!15}
40 & 0.00029808 &
\(0.0025153236949469683\Big((x_{35}-(x_{45}-x_{18}))\big(x_{43}-0.1117044718447674\,x_{24}x_{15}x_{10}\big)
+ (x_{46}-x_{22})x_{47}
+ \big(x_{45}-0.4030994964706104(x_{51}-x_{46}-2.472476686808669-x_{21})-x_{15}\big)^2\Big)
+ 0.3125372513473696\) \\

42 & 0.00029334 &
\(0.002559798232109027\Big((x_{35}-(x_{45}-x_{18}))\big(x_{43}-0.10673373507090549\,x_{10}x_{24}x_{15}\big)
+ \big(x_{45}-0.33261097044161836(x_{51}-x_{46}+x_{7}-x_{21})-x_{15}\big)^2
+ x_{47}(x_{46}-x_{22}) + x_{51}\Big)
+ 0.306355949551514\) \\

44 & 0.00027908 &
\(0.002490439657321843\Big((x_{46}-x_{22})x_{47}
+ (x_{35}-(x_{45}-x_{18}))\big(x_{43}-0.12624988881508498\,x_{15}x_{24}x_{10}\big)
+ \big(x_{45}-0.4440134981610858(x_{51}-0.5187357361284024(x_{46}-x_{23}-x_{7})-x_{21})-x_{15}\big)^2\Big)
+ 0.31244325316937577\) \\

46 & 0.00026917 &
\(0.0024897226018737156\Big(x_{10}
+ (x_{15}-(x_{45}-0.47724653358206776(x_{51}-0.4999848079129539(x_{46}-x_{23}-x_{7})-x_{21})))^2
+ (x_{35}-(x_{45}-x_{18}))\big(x_{43}-0.1344127470430672\,x_{15}x_{24}x_{10}\big)
+ (x_{46}-x_{22})x_{47}\Big)
+ 0.30744355777844795\) \\

48 & 0.00025901 &
\(0.0025884970138480242\Big((x_{46}-x_{22})x_{47}
+ \big(x_{45}-0.455787070592263(x_{51}-0.4805310670016915(x_{46}-x_{23}-x_{7})-x_{21})-x_{15}\big)^2
+ (x_{35}-(x_{45}-x_{18}))\big(x_{43}-0.1367129133385923\,x_{24}x_{10}x_{15}\big)
+ (x_{10}+x_{24})\Big)
+ 0.3013569751203911\) \\

50 & 0.00025835 &
\(0.0025958429222822923\Big(((x_{46}-x_{22})+0.34991507947644734\,x_{10})x_{47}
+ (x_{35}-(x_{45}-x_{18}))\big(x_{43}-0.13643454842977537\,x_{10}x_{24}x_{15}\big)
+ x_{24}
+ \big(x_{15}-(x_{45}-0.4396520838509348(x_{51}-0.4892197777096199(x_{46}-x_{23}-x_{7})-x_{21}))\big)^2\Big)
+ 0.3031685893578367\) \\
\end{longtable}

\clearpage

\section{Decision Tree Illustrations} \label{app::decision-tree-pictures}
\begin{center}
    \begin{figure}[h]
        \centering
        \includegraphics[width=\linewidth]{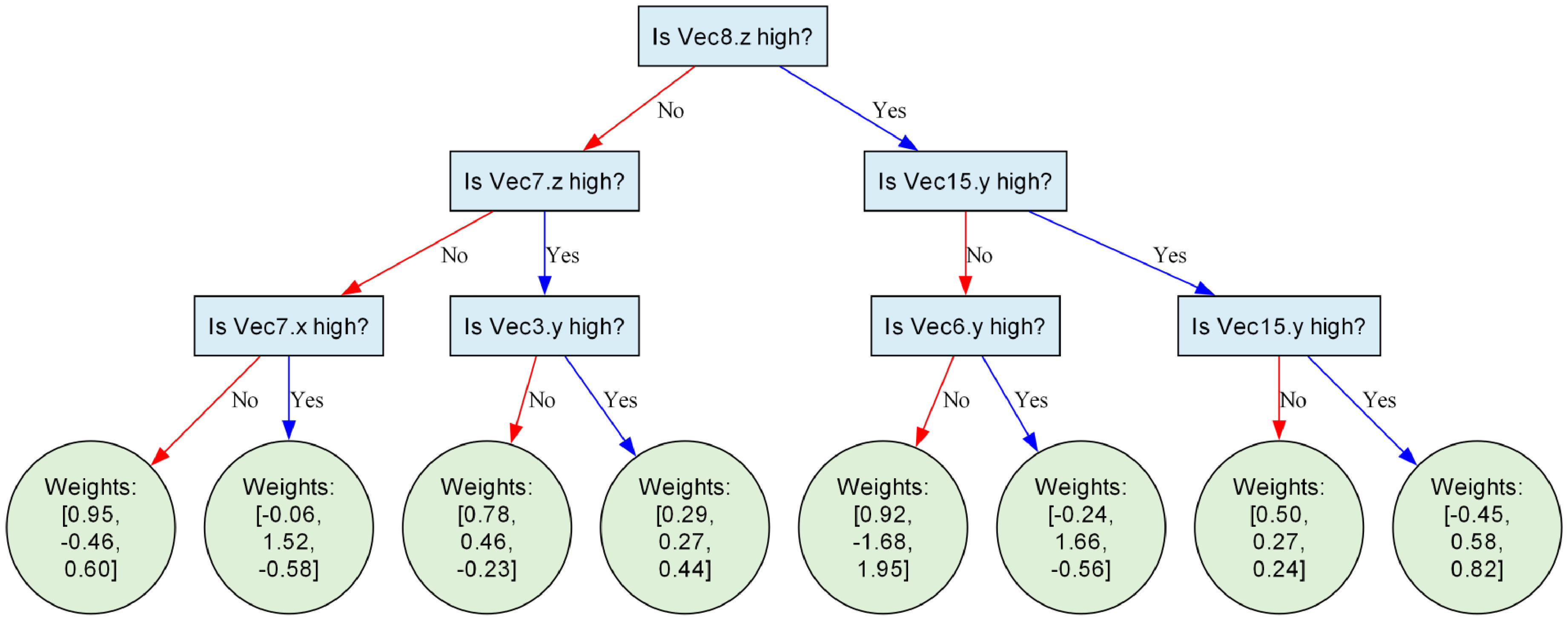}
        \caption{Depth 3 soft-decision tree}
    \end{figure}
\end{center}

\begin{center}
    \begin{figure}[h]
        \centering
        \includegraphics[width=1.1\linewidth]{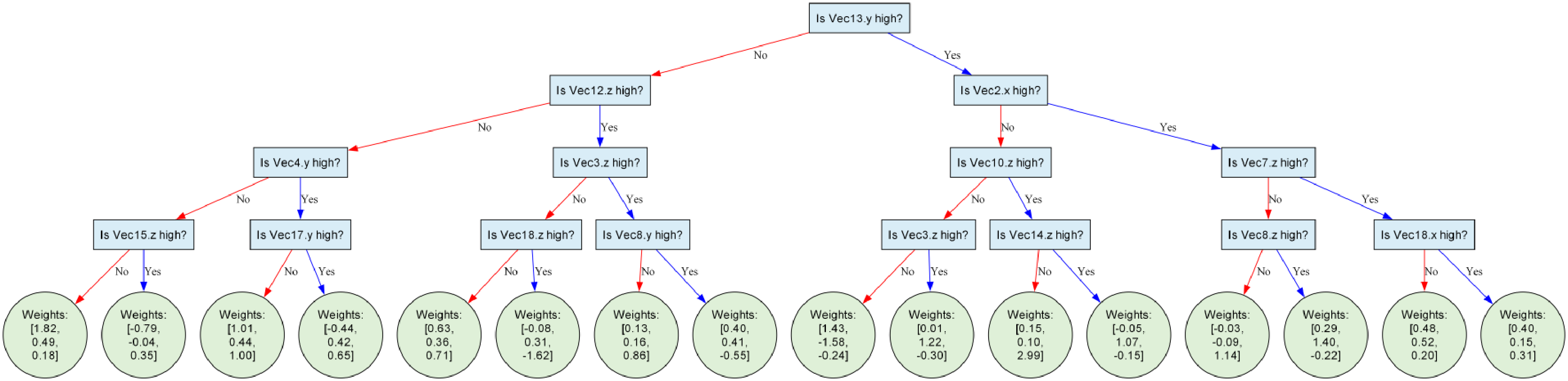}
        \caption{Depth 4 soft-decision tree}
        \label{fig:placeholder}
    \end{figure}
\end{center}

\begin{center}
    \begin{figure}[h]
        \centering
        \includegraphics[width=\linewidth]{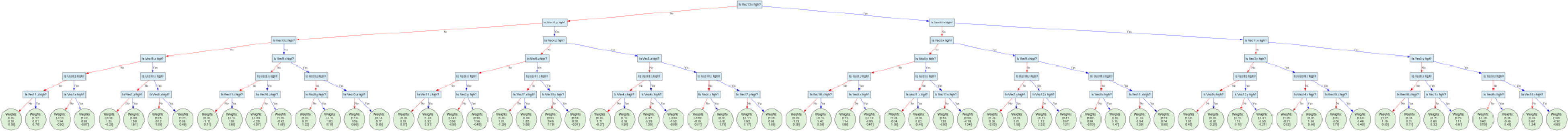}
        \caption{Depth 6 soft-decision tree}
        \label{fig:placeholder}
    \end{figure}
\end{center}

\clearpage

\section{Polynomial Support Sets for Experiments}
\label{app:support-sets}

\subsection{Relative pose with unknown focal length}
\label{app:relpose-support}
Ordering variables as $(l_1,l_2,p)$, the ten polynomials have support set (i.e.\ the exponent vectors of the monomials in $X$):
\begin{equation}
\label{eq:relpose-support-set}
\begin{aligned}
\mathcal{A}=\Big\{&
(3,0,2),(3,0,1),(3,0,0),
(2,1,2),(1,2,2),(0,3,2),
(2,1,1),(1,2,1),(0,3,1),
(2,1,0),(1,1,2), \\ 
& 
(2,0,2),
(0,2,2),(1,0,3), (1,2,0),
(0,3,0), (2,0,1),(0,1,3),
(1,1,1),(0,2,1), (1,0,2),
(0,1,2), \\
&
(0,0,3), (2,0,0),
(1,1,0),(0,2,0), (1,0,1),
(0,1,1),(0,0,2), (1,0,0),
(0,1,0),(0,0,1),(0,0,0)
\Big\}.
\end{aligned}
\end{equation}

\subsection{Three-view triangulation}
\label{app:triangulation-support}
Ordering variables as $(x,y,z)$, the three polynomials have support sets:
\begin{equation}
\label{eq:triangulation-support-sets}
\begin{aligned}
\mathcal{A}_1=\Big\{&
(4,2,0),(4,0,2),(3,3,0),\;
(3,2,1),(3,2,0),(3,1,2), 
(3,0,3),(3,0,2),(1,3,2),\\
&
(1,2,3),(1,2,2),
(0,4,2), (0,3,3),(0,3,2),(0,2,4),\;
(0,2,3),(0,2,2)
\Big\},\\[0.5em]
\mathcal{A}_2=\Big\{&
(4,0,2),(3,3,0),(3,1,2),\;
(3,0,3),(3,0,2),(2,4,0),
(2,3,1),(2,3,0), \\ 
& 
(2,1,3), (2,1,2), (2,0,4),(2,0,3),
(2,0,2),(1,3,2),(0,4,2),\;
(0,3,3),(0,3,2)
\Big\},\\[0.5em]
\mathcal{A}_3=\Big\{&
(4,2,0),(3,3,0),(3,2,1),\;
(3,2,0),(3,0,3),(2,4,0),
(2,3,1), (2,3,0), \\ 
& 
(2,2,1), (2,2,0),(2,1,3),(2,0,4),
(2,0,3),(1,2,3),(0,3,3),\;
(0,2,4),(0,2,3)
\Big\}.
\end{aligned}
\end{equation}

\subsection{\(n\)-site phosphorylation system (\(n=14\))}
\label{app:nsite-support}
Ordering variables as \((s_0,e)\), both polynomials share the same support set:
\begin{equation}
\label{eq:nsite-support-set}
\begin{aligned}
\mathcal{A}=\Big\{&
(1,0),(0,1),
(1,1),(1,2),(1,3),(1,4),(1,5),(1,6),(1,7),\\
&
(1,8),(1,9),(1,10),(1,11),(1,12),(1,13),(1,14),
(0,0)
\Big\}.
\end{aligned}
\end{equation}

\subsection{Wnt shuttle model system}
\label{app:wnt-support}
Ordering variables as \((x_1, x_2, \ldots, x_{19})\), the twenty-four polynomials have support sets:
\begin{equation}
\label{eq:wnt-support-set}
\begin{aligned}
\mathcal{A}_1=\Big\{&
(1,0,0,0,0,0,0,0,0,0,0,0,0,0,0,0,0,0,0),\;
(0,1,0,0,0,0,0,0,0,0,0,0,0,0,0,0,0,0,0)
\Big\},\\[0.4em]
\mathcal{A}_2=\Big\{&
(1,0,0,0,0,0,0,0,0,0,0,0,0,0,0,0,0,0,0),\;
(0,1,0,0,0,0,0,0,0,0,0,0,0,0,0,0,0,0,0),\\
&
(0,0,1,0,0,0,0,0,0,0,0,0,0,0,0,0,0,0,0),\;
(0,1,0,1,0,0,0,0,0,0,0,0,0,0,0,0,0,0,0),\\
&
(0,0,0,0,0,0,0,0,0,0,0,0,0,1,0,0,0,0,0)
\Big\},\\[0.4em]
\mathcal{A}_3=\Big\{&
(0,1,0,0,0,0,0,0,0,0,0,0,0,0,0,0,0,0,0),\;
(0,0,1,0,0,0,0,0,0,0,0,0,0,0,0,0,0,0,0),\\
&
(0,0,1,0,0,1,0,0,0,0,0,0,0,0,0,0,0,0,0),\;
(0,0,0,0,0,0,0,0,0,0,0,0,0,0,1,0,0,0,0)
\Big\},\\[0.4em]
\mathcal{A}_4=\Big\{&
(0,1,0,1,0,0,0,0,0,0,0,0,0,0,0,0,0,0,0),\;
(0,0,0,1,0,0,0,0,0,1,0,0,0,0,0,0,0,0,0),\\
&
(0,0,0,0,0,0,0,0,0,0,0,0,0,1,0,0,0,0,0),\;
(0,0,0,0,0,0,0,0,0,0,0,0,0,0,0,1,0,0,0),\\
&
(0,0,0,0,0,0,0,0,0,0,0,0,0,0,0,0,0,1,0)
\Big\},\\[0.4em]
\mathcal{A}_5=\Big\{&
(0,0,0,0,1,0,0,0,0,0,0,0,0,0,0,0,0,0,0),\;
(0,0,0,0,0,0,1,0,0,0,0,0,0,0,0,0,0,0,0),\\
&
(0,0,0,0,1,0,0,1,0,0,0,0,0,0,0,0,0,0,0),\;
(0,0,0,0,0,0,0,0,0,0,0,0,0,1,0,0,0,0,0),\\
&
(0,0,0,0,0,0,0,0,0,0,0,0,0,0,0,1,0,0,0)
\Big\},\\[0.4em]
\mathcal{A}_6=\Big\{&
(0,0,1,0,0,1,0,0,0,0,0,0,0,0,0,0,0,0,0),\;
(0,0,0,0,0,1,0,0,0,0,1,0,0,0,0,0,0,0,0),\\
&
(0,0,0,0,0,0,0,0,0,0,0,0,0,0,1,0,0,0,0),\;
(0,0,0,0,0,0,0,0,0,0,0,0,0,0,0,0,1,0,0),\\
&
(0,0,0,0,0,0,0,0,0,0,0,0,0,0,0,0,0,0,1)
\Big\},\\[0.4em]
\mathcal{A}_7=\Big\{&
(0,0,0,0,1,0,0,0,0,0,0,0,0,0,0,0,0,0,0),\;
(0,0,0,0,0,0,1,0,0,0,0,0,0,0,0,0,0,0,0),\\
&
(0,0,0,0,0,0,1,0,1,0,0,0,0,0,0,0,0,0,0),\;
(0,0,0,0,0,0,0,0,0,0,0,0,0,0,1,0,0,0,0),\\
&
(0,0,0,0,0,0,0,0,0,0,0,0,0,0,0,0,1,0,0)
\Big\},\\[0.4em]
\mathcal{A}_8=\Big\{&
(0,0,0,0,1,0,0,1,0,0,0,0,0,0,0,0,0,0,0),\;
(0,0,0,0,0,0,0,0,0,0,0,0,0,0,0,1,0,0,0)
\Big\},\\[0.4em]
\mathcal{A}_9=\Big\{&
(0,0,0,0,0,0,1,0,1,0,0,0,0,0,0,0,0,0,0),\;
(0,0,0,0,0,0,0,0,0,0,0,0,0,0,0,0,1,0,0)
\Big\},\\[0.4em]
\mathcal{A}_{10}=\Big\{&
(0,0,0,0,0,0,0,0,0,0,0,0,0,0,0,0,0,0,0),\;
(0,0,0,0,0,0,0,0,0,1,0,0,0,0,0,0,0,0,0),\\
&
(0,0,0,1,0,0,0,0,0,1,0,0,0,0,0,0,0,0,0),\;
(0,0,0,0,0,0,0,0,0,0,1,0,0,0,0,0,0,0,0),\\
&
(0,0,0,0,0,0,0,0,0,0,0,0,0,0,0,0,0,1,0)
\Big\},
\end{aligned}
\end{equation}

\begin{equation}
\begin{aligned}
\mathcal{A}_{11}=\Big\{&
(0,0,0,0,0,0,0,0,0,0,1,0,0,0,0,0,0,0,0),\;
(0,0,0,0,0,0,0,0,0,1,0,0,0,0,0,0,0,0,0),\\
&
(0,0,0,0,0,1,0,0,0,0,1,0,0,0,0,0,0,0,0),\;
(0,0,0,0,0,0,0,0,0,0,1,1,0,0,0,0,0,0,0),\\
&
(0,0,0,0,0,0,0,0,0,0,0,0,1,0,0,0,0,0,0),\;
(0,0,0,0,0,0,0,0,0,0,0,0,0,0,0,0,0,0,1)
\Big\},\\[0.4em]
\mathcal{A}_{12}=\Big\{&
(0,0,0,0,0,0,0,0,0,0,1,1,0,0,0,0,0,0,0),\;
(0,0,0,0,0,0,0,0,0,0,0,0,1,0,0,0,0,0,0)
\Big\},\\[0.4em]
\mathcal{A}_{13}=\Big\{&
(0,0,0,0,0,0,0,0,0,0,1,1,0,0,0,0,0,0,0),\;
(0,0,0,0,0,0,0,0,0,0,0,0,1,0,0,0,0,0,0)
\Big\}, \\[0.4em]
\mathcal{A}_{14}=\Big\{&
(0,1,0,1,0,0,0,0,0,0,0,0,0,0,0,0,0,0,0),\;
(0,0,0,0,0,0,0,0,0,0,0,0,0,1,0,0,0,0,0)
\Big\},\\[0.4em]
\mathcal{A}_{15}=\Big\{&
(0,0,1,0,0,1,0,0,0,0,0,0,0,0,0,0,0,0,0),\;
(0,0,0,0,0,0,0,0,0,0,0,0,0,0,1,0,0,0,0)
\Big\},\\[0.4em]
\mathcal{A}_{16}=\Big\{&
(0,0,0,0,1,0,0,1,0,0,0,0,0,0,0,0,0,0,0),\;
(0,0,0,0,0,0,0,0,0,0,0,0,0,0,0,1,0,0,0)
\Big\},\\[0.4em]
\mathcal{A}_{17}=\Big\{&
(0,0,0,0,0,0,1,0,1,0,0,0,0,0,0,0,0,0,0),\;
(0,0,0,0,0,0,0,0,0,0,0,0,0,0,0,0,1,0,0)
\Big\},\\[0.4em]
\mathcal{A}_{18}=\Big\{&
(0,0,0,1,0,0,0,0,0,1,0,0,0,0,0,0,0,0,0),\;
(0,0,0,0,0,0,0,0,0,0,0,0,0,0,0,0,0,1,0)
\Big\},\\[0.4em]
\mathcal{A}_{19}=\Big\{&
(0,0,0,0,0,1,0,0,0,0,1,0,0,0,0,0,0,0,0),\;
(0,0,0,0,0,0,0,0,0,0,0,0,0,0,0,0,0,0,1)
\Big\},\\[0.4em]
\mathcal{A}_{20}=\Big\{&
(0,0,0,0,0,0,0,0,0,0,0,0,0,0,0,0,0,0,0),\;
(1,0,0,0,0,0,0,0,0,0,0,0,0,0,0,0,0,0,0),\\
&
(0,1,0,0,0,0,0,0,0,0,0,0,0,0,0,0,0,0,0),\;
(0,0,1,0,0,0,0,0,0,0,0,0,0,0,0,0,0,0,0),\\
&
(0,0,0,0,0,0,0,0,0,0,0,0,0,1,0,0,0,0,0),\;
(0,0,0,0,0,0,0,0,0,0,0,0,0,0,1,0,0,0,0)
\Big\},\\[0.4em]
\mathcal{A}_{21}=\Big\{&
(0,0,0,0,0,0,0,0,0,0,0,0,0,0,0,0,0,0,0),\;
(0,0,0,1,0,0,0,0,0,0,0,0,0,0,0,0,0,0,0),\\
&
(0,0,0,0,1,0,0,0,0,0,0,0,0,0,0,0,0,0,0),\;
(0,0,0,0,0,1,0,0,0,0,0,0,0,0,0,0,0,0,0),\\
&
(0,0,0,0,0,0,1,0,0,0,0,0,0,0,0,0,0,0,0),\;
(0,0,0,0,0,0,0,0,0,0,0,0,0,1,0,0,0,0,0),\\
&
(0,0,0,0,0,0,0,0,0,0,0,0,0,0,1,0,0,0,0),\;
(0,0,0,0,0,0,0,0,0,0,0,0,0,0,0,1,0,0,0),\\
&
(0,0,0,0,0,0,0,0,0,0,0,0,0,0,0,0,1,0,0),\;
(0,0,0,0,0,0,0,0,0,0,0,0,0,0,0,0,0,1,0),\\
&
(0,0,0,0,0,0,0,0,0,0,0,0,0,0,0,0,0,0,1)
\Big\},\\[0.4em]
\mathcal{A}_{22}=\Big\{&
(0,0,0,0,0,0,0,0,0,0,0,0,0,0,0,0,0,0,0),\;
(0,0,0,0,0,0,0,1,0,0,0,0,0,0,0,0,0,0,0),\\
&
(0,0,0,0,0,0,0,0,0,0,0,0,0,0,0,1,0,0,0)
\Big\},\\[0.4em]
\mathcal{A}_{23}=\Big\{&
(0,0,0,0,0,0,0,0,0,0,0,0,0,0,0,0,0,0,0),\;
(0,0,0,0,0,0,0,0,1,0,0,0,0,0,0,0,0,0,0),\\
&
(0,0,0,0,0,0,0,0,0,0,0,0,0,0,0,0,1,0,0)
\Big\},
\end{aligned}
\tag{7}
\end{equation}

\begin{equation}
\begin{aligned}
\mathcal{A}_{24}=\Big\{&
(0,0,0,0,0,0,0,0,0,0,0,0,0,0,0,0,0,0,0),\;
(0,0,0,0,0,0,0,0,0,0,0,1,0,0,0,0,0,0,0),\\
&
(0,0,0,0,0,0,0,0,0,0,0,0,1,0,0,0,0,0,0)
\Big\}.
\end{aligned}
\tag{7}
\end{equation}

\end{document}